\documentclass[twocolumn]{emulateapj}  
\usepackage{times}

\usepackage{graphicx,bm,amssymb}
\usepackage{epsfig}
\usepackage{amssymb}
\usepackage{natbib}
\usepackage{pstricks}
\usepackage{psfrag}
\usepackage[colorlinks=true,linkcolor=blue,citecolor=blue]{hyperref}
\usepackage{float}
\usepackage[caption = false]{subfig}
\def\araa{ARA\&A}
\def\apj{ApJ}
\def\apjl{ApJ}
\def\apjs{ApJS}
\def\apss{Ap\&SS}
\def\aap{A\&A}
\def\aaps{A\&AS}
\def\mnras{MNRAS}
\def\prd{Phys.~Rev.~D}
\def\nat{Nature}
\def\physrep{Phys.~Rep.}

\newcommand{\be}{\begin{equation}}
\newcommand{\ee}{\end{equation}}
\newcommand{\bary}{\begin{eqnarray}}
\newcommand{\eary}{\end{eqnarray}}

\shorttitle{GRB 170817A}
\shortauthors{Fraija N. et al.}

\begin{document}
\title{The short GRB 170817A: Modelling the off-axis emission and\\ implications on the ejecta magnetization}





\author{N. Fraija$^{1\dagger}$, F. De Colle$^2$, P. Veres$^3$, S. Dichiara$^{1,4,5}$,  R. Barniol Duran$^6$,   A. Galvan-Gamez$^{1}$ and A.C. Caligula do E. S. Pedreira$^{1,7}$} 
%
\affil{$^1$ Instituto de Astronom\' ia, Universidad Nacional Aut\'onoma de M\'exico, Circuito Exterior, C.U., A. Postal 70-264, 04510 Cd. de M\'exico,  M\'exico\\
$^2$ Instituto de Ciencias Nucleares, Universidad Nacional Aut\'onoma de M\'exico, Circuito Exterior, C.U., A. Postal 70-264, 04510 Cd. de M\'exico,  M\'exico\\
$^3$ Center for Space Plasma and Aeronomic Research (CSPAR), University of Alabama in Huntsville, Huntsville, AL 35899, USA\\
$^4$ Department of Astronomy, University of Maryland, College Park, MD 20742-4111, USA\\
$^5$ Astrophysics Science Division, NASA Goddard Space Flight Center, 8800 Greenbelt Rd, Greenbelt, MD 20771, USA\\
$^6$ Department of Physics and Astronomy, California State University, Sacramento, 6000 J Street, Sacramento, CA 95819-6041, USA\\
$^7$ Instituto de Matem\'atica, Estat\'isca e F\'isica, Universidade Federal do Rio Grande, Rio Grande 96203-900, Brasil\\
}
\email{$\dagger$nifraija@astro.unam.mx}
%
\begin{abstract}
The short GRB 170817A, detected  by the Fermi Gamma-ray Burst Monitor, orbiting satellites and ground-based telescopes, was the electromagnetic counterpart of a gravitational-wave transient (GW170817) from a binary neutron star merger.  After this merger the $\gamma$-ray light curve exhibited a faint peak at $\sim$ 1.7s and  the X-ray, optical and radio light curves displayed an extended emission which increased in brightness up to $\sim$ 160 days.  In this paper,  we  show that the X-ray, optical and radio fluxes are consistent with the synchrotron forward-shock model viewed off-axis when the matter in the outflow is  parametrized through  a power law velocity distribution.   We discuss the origin of the $\gamma$-ray peak in terms of internal and external shocks.  We show that the $\gamma$-ray flux might be consistent with a synchrotron self-Compton reverse-shock model observed at high latitudes.   Comparing the best-fit values obtained after describing the $\gamma$-ray,  X-ray, optical and radio fluxes with our model,  we find that the afterglow and $\gamma$-ray emission occurred in different regions and also evidence to propose that the progenitor environment was entrained with magnetic fields and therefore, we argue for the presence of the magnetic field amplification in the binary neutron star merger.  
\end{abstract}
\keywords{gamma-rays bursts: individual (GRB 170817A) ---  Physical data and processes: acceleration of particles  --- Physical data and processes: radiation mechanism: nonthermal --- ISM: general - magnetic fields}
\section{Introduction}
One of the most fascinating extragalactic events are gamma-ray bursts (GRBs).   They are known to exhibit a vast variety of spectral and temporal properties. Based on the standard GRB durations and spectral hardness two kind of progenitor populations have been amply accepted, short ($T_{90}<$ 2 s) and long ($T_{90}>$ 2 s) GRBs \citep[for review, see][]{2004IJMPA..19.2385Z, 2015PhR...561....1K}.    Although the discoveries and subsequent studies of long GRBs (lGRBs) have been marked by many successes, the study of short GRBs (sGRBs) has proven to be much more challenging.   Significant advances in sGRBs were achieved with the discovery of the first host galaxies and the  observations of  multiwavelength afterglows \citep[for reviews, see][]{2007PhR...442..166N, 2014ARA&A..52...43B}.  Several lines of evidence have associated the sGRB progenitors with  the merger of compact object binaries comprised of a neutron star binary (NS-NS) or a neutron star - black hole (NS-BH) \citep{1989Natur.340..126E, 1992ApJ...395L..83N, 2007NJPh....9...17L, 2004ApJ...608L...5L,2005ApJ...632..421L, 2007PhR...442..166N}.  These progenitors are promising candidates to release gravitational waves (GWs) accompanied by an isotropic optical/infrared  counterpart, the so-called kilonova or macronova \citep{1998ApJ...507L..59L, 2005ApJ...634.1202R, 2010MNRAS.406.2650M, 2013ApJ...774...25K, 2017LRR....20....3M}. Because of neutron-rich ejecta from these progenitors, a  kilonova/macronova is produced via radioactive decay of unstable heavy nuclei created in the rapid neutron capture (r-process) nucleosynthesis \citep{1974ApJ...192L.145L, 1976ApJ...210..549L}. In addition, a cocoon emission and a delayed non-thermal radiation in radio wavelengths, originated from the interaction of the merger ejecta with the circumburst medium, are expected from these events \citep{2011Natur.478...82N, 2013MNRAS.430.2121P, 2015MNRAS.450.1430H, 2017MNRAS.471.1652L, 2017ApJ...848L...6L}.    On the other hand, using two non-spinning magnetized NSs initially separated by 48 km with 1.4 solar masses, \cite{2006Sci...312..719P} presented through simulations the magnetic field evolution in a binary NS merger.  The main result is that the corresponding magnetic field, of $\sim 10^{12}$ G, present in a NS can be dramatically amplified by several orders of magnitude after the merger.  The magnetic field strength that can be reached during the first milliseconds through Kevin-Helmholtz instabilities and turbulent amplification is much higher than $\sim10^{15}$ G \citep{2006Sci...312..719P, 2013ApJ...769L..29Z, 2009MNRAS.399L.164G,2010A&A...515A..30O}. Therefore, a degree of magnetization in the ejecta could be expected in the binary NS merger.\\
During the last decade,  the observation of optical and gamma-ray polarization \citep[e.g. see][]{2009Natur.462..767S, 2007Sci...315.1822M, 2017Natur.547..425T, 2013Natur.504..119M}  and the modelling of $\gamma$-ray, X-ray and optical bright peaks  which suggests a stronger magnetic field in the reverse-shock region than in the forward-shock region \citep{2003ApJ...595..950Z, 2003ApJ...597..455K, 2000ApJ...545..807K,  2007ApJ...655..391K, 2015ApJ...804..105F, 2016ApJ...831...22F,2012ApJ...751...33F, 2017ApJ...848...94F} have provided overwhelming evidence that some lGRB progenitors are endowed of intense magnetic fields \citep[see e.g.][and references therein]{1992Natur.357..472U}. In the context of sGRBs,  \cite{2016ApJ...831...22F} proposed that the bright peak exhibited at the prompt/early-afterglow phase could be correlated with the degree of magnetization present in the jet.  In particular, the bright peak presented in the large area telescope (LAT) light curve and interpreted in the reverse-shock context indicated a compelling evidence that the central engine in GRB 090510 was magnetized, being the  magnetic field amplification in the binary NS merger the most promising candidate.\\
%
%
\\
%
%
%
On the other hand, the transition between the prompt emission and the afterglow is one of the most interesting and least understood  phases.  The prompt decay phase is attributed to emission from regions located at high latitudes, i.e. from regions located at viewing angles ($\theta_{\rm obs}$) larger by at least a factor  $\theta_j\sim\frac{1}{\Gamma}$ with respect to the line of sight (the curvature effect or high-latitude emission).   When this effect is present, after the gamma-ray emission from the observer's line of sight has ceased, the off-axis flux at $\theta_{\rm obs}>\theta_j$ is dramatically suppressed unless the burst is very luminous or viewed from near its edge.  Because of the curvature effect, the onset of the afterglow could be overlapped with the high-latitude emission.  Radiation generated at the reverse shock would decay fast due to the angular time delay effect \citep{2000ApJ...541L..51K}.   Once the quickly decaying high-latitude emission is small enough the afterglow emission can be observed \citep{2000ApJ...537..785D, 2002ApJ...570L..61G, 1999A&AS..138..491R}.\\ 
%
%
GRB 170817A, the electromagnetic counterpart of the gravitational-wave transient associated with a NS-NS coalescence  \citep[GW170817;][]{PhysRevLett.119.161101,2041-8205-848-2-L12}, was detected by the Gamma-ray Burst Monitor (GBM) onboard Fermi Gamma-ray Space Telescope at 12:41:06 UTC, 2017 August 17 \citep{2017arXiv171005446G}.  Promptly, this burst was monitored in several electromagnetic bands by multiple ground-based telescopes and satellites \citep[see e.g.][and references therein]{2041-8205-848-2-L12}.    By considering the low luminosity observed in GRB 170817A  the $\gamma$-ray flux has been associated to different emission mechanisms \citep{2017arXiv171005896G, 2017arXiv171005897B, 2017arXiv171100243K, 2018arXiv180302978F, 2017arXiv170807488K, 2018arXiv180905099K} and the X-ray, optical and radio afterglow with synchrotron forward-shock models when the relativistic jet viewed off-axis and/or cocoon are decelerated in an homogeneous low density medium $\sim 10^{-5}$ - $10^{-2}\,{\rm cm^{-3}}$ \citep{2017ApJ...848L..34M, 2017arXiv171005905I, 2017arXiv171111573M, 2017arXiv171203237L, 2017arXiv171006421G, 2017ApJ...848L..21A, 2017ApJ...848L..20M, 2017Sci...358.1559K, 2017arXiv171005822P, 2018ApJ...853L..13W, 2017arXiv171006426G, 2017arXiv171006421G}.\\
%
%
In this paper, we present a comprehensive analysis and description of the short GRB 170817A, in the context of an off-axis jet, when the matter in the outflow is  parametrized through  a power law velocity distribution.  The paper is arranged as follows: In Section 2, a brief description of the multiwavelength observations and GBM data reduction is presented.  In Section 3,  we model the non-thermal multiwavelength observations in GRB 170817A and discuss the implications on the ejecta magnetization and conclusions are given in Section 4.\\
\vspace{2cm}
\section{GRB 170817A}
\subsection{Multiwavelength upper limits and observations}
GRB 170817A was detected by the GBM-Fermi Telescope at 12:41:06 UTC, 2017 August 17 \citep{2017GCN.21520....1V, 2017arXiv171005446G}.  This detection was consistent with a gravitational-wave transient observed by LIGO and Virgo observatories. This observational transient was associated with a NS-NS coalescence with merger time 12:41:04 UTC $\sim$ 2 s  before the GBM trigger \citep{PhysRevLett.119.161101, 2041-8205-848-2-L12}. Immediately afterwards, an exhaustive  multiwavelength campaign was launched in order to look for an isotropic  electromagnetic counterpart in the optical and infrared bands  \citep[see e.g.][and references therein]{2017arXiv171005452C}.  A bright transient in the optical i-band  with magnitude $m_i=17.057\pm 0.0018$ was observed by the 1-meter Swope telescope at Las Campanas Observatory in Chile at 10.87 hours (August 17 at 23:33 UTC) after the GMB trigger and  afterward during the following 12 hours  by multiple ground-based and orbiting optical/IR  telescopes. In addition, linear polarization in optical bands was reported, revealing the geometry of the emitting region.  This transient  was located coming from the center of the galaxy NGC 4993 at a distance of 40 Mpc.\\ 
Distinct X-ray observations were carried out by several orbiting satellites during the following 8 days without any detection but providing constraining limits \citep[i.e. see][]{2017ApJ...848L..20M}.  From the 9th up to 256th day  after the merger, X-ray detections have been reported by Chandra and XMM-Newton observatories  \citep{troja2017a, 2018arXiv180103531M, 2018arXiv180502870A, 2018arXiv180106164D}.  Optical observations and upper limits collected  with the Advanced  Camera  for  Surveys  Wide  Field  Camera  aboard on  the Hubble Satellite Telescope (HST), have been performed since $\sim$ 100 days after the trigger \citep{2018arXiv180102669L, 2018arXiv180103531M, 2018arXiv180502870A}.  On the sixteenth day after the post-trigger and for more than seven months, the radio counterpart at 3 and 6 GHz was obtained by Very Large Array \citep[VLA; ][]{2041-8205-848-2-L12, 2017arXiv171111573M, 2018ApJ...858L..15D, 2017arXiv171005435H}.\\ 
\subsection{GBM data reduction}
Event data files were obtained using the GBM trigger time for GRB 170817A 04:47:43 UT on 2017 August 17 \citep{2017GCN.21520....1V, 2013ApJS..209...11A}. Fermi-GBM data in the energy range of 10 - 1000 keV were reduced using the public database at the Fermi Website\footnote{\url{http://fermi.gsfc.nasa.gov/ssc/data}} and the position of this burst is found to be at the coordinates (J2000) RA = 176$^\circ$.8, DEC = -39$^\circ$.8, with an error circle of radius 11.6$^{\circ}$. No other sources in the LAT catalog or background emission are considered due to the duration of the event.\\
 Flux values are derived using the spectral analysis package RMfit, version 432\footnote{\url{https://fermi.gsfc.nasa.gov/ssc/data/analysis/rmfit/}}.  To analyze the signal we use the time-tagged event (TTE) files of the three triggered NaI detectors ${\rm n_1}$, ${\rm n_2}$ and ${\rm n_5}$. Different spectral models are used to fit the spectrum over different duration periods. Each time bin is chosen adopting a trade-off between the minimum signal needed to derive a spectrum and the minimum resolution required to preserve the shape of the time evolution. The Comptonized (a power law with exponential cutoff, hereafter referred as CPL) and the simple power-law (PL) functions are used to fit the spectrum up to 0.448 s around the GBM trigger time.  The spectral analysis during the time interval [-0.320 s, 0.448 s] after the trigger is reported in Table \ref{table1:gbm_analysis}.  This table shows the time interval (column one), spectral model (column 2), spectral index (column three),  energy peak (column four), temperature of black body (BB) function and the C-Stat/dof test (last column).    After the 0.512 s the spectrum fits better using a BB model.\\
\vspace{0.5cm}
\section{Description of the non-thermal multiwavelength observations in GRB 170817A}\label{sec3}
\subsection{Modeling the $\gamma$-ray flux}
\subsubsection{Light curve analysis and description}
\begin{figure}
{ \centering
\resizebox*{0.45\textwidth}{0.26\textheight}
{\includegraphics{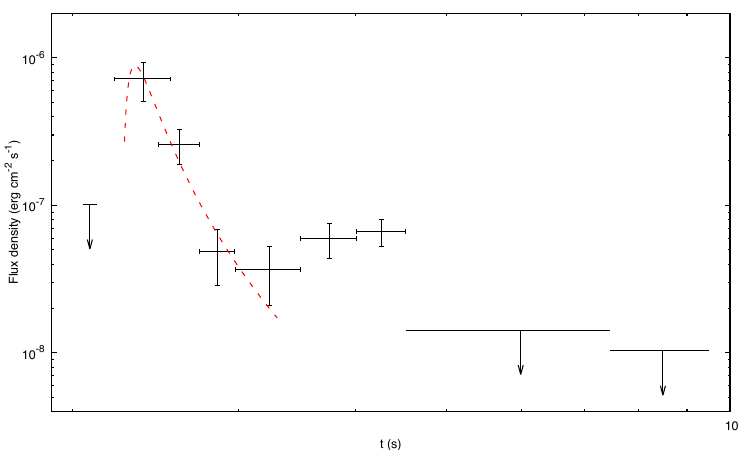}}
}
\caption{GBM light curve and upper limits in the energy range of 10 - 1000 keV of GRB 170817A. The red line corresponds to the best-fit curve using  a function $F(t) \propto \left(\frac{t-t_0}{t_0}\right)^{-\alpha}\,e^{-\frac{\tau}{t-t_0}}$ \citep{2006Natur.442..172V}. }
\label{fig1:gbm}
\end{figure}

Figure \ref{fig1:gbm} shows the GBM light curve and upper limits  in the energy range of 10 - 1000 keV, although no significant flux was observed above 300 keV.  The CPL function showed a cutoff energy of 185 keV and  the corresponding isotropic energy obtained was $E_{\rm \gamma, iso}\simeq5\times 10^{46}$ erg, with $T_{90}$= 2 s  \citep{2017GCN.21520....1V}.  The  GBM light curve exhibited a peak around $\sim$ 1.7 s after the gravitational-wave trigger, followed by a fast decay.  The Chi-square ($\chi^2$) minimization method, developed in the ROOT software package \citep{1997NIMPA.389...81B}, was used in order to fit the GBM light curve with the function: $F_\nu(t)\propto (\frac{t-t_0}{t_0})^{-\alpha_\gamma} e^{-\frac{\tau}{t-t_0}}$ \citep{2006Natur.442..172V} where $t_0$ is the starting time, $\tau$ is the timescale of the flux rise and $\alpha_\gamma$  the power index of fast decay.  The best-fit values of parameters are reported in Table \ref{table2: fit_gbm}.\\
\\
We derive the spectral parameters of GBM data for different time intervals, as shown in Table \ref{table1:gbm_analysis}.   Two different time intervals, starting from -0.320 s,  were used to fit with a CPL function. 
The best-fit value of the spectral index for the interval  [-0.320 s, 0.256 s] was $-0.955\pm 0.309$.  The remaining time interval was divided in two and analyzed with a PL function.  For the interval  [0.256 s, 0320 s] a spectral index of  $-1.749\pm 0.434$ was obtained and for interval  [0.320 s, 0.448 s] the corresponding spectral index was  $-2.150\pm 0.472$.  The spectral fit parameters associated with the  $\gamma$-ray peak reveal  a hard-to-soft spectral evolution.\\
\\
\cite{2018arXiv180207328V} analyzed, in the GBM data, the evolution of the peak energy with a CPL model.  Using  a simple PL $E_{\rm peak} \propto (t-t_{\rm shift})^{\rm -q}$ to model the decay phase, they obtained the best-fit value of  ${\rm q}=0.97\pm 0.35$ for $t_{\rm shift}=-0.15\pm 0.04\, {\rm s}$.\\
\\
Based on the best-fit values obtained from the analysis of the GBM data and reported in Tables \ref{table1:gbm_analysis} and \ref{table2: fit_gbm},  we discuss the origin of the $\gamma$-ray light curve in terms of internal and external shocks.
\paragraph {1.  The $\gamma$-ray peak  $\delta t_{\rm var}/ T_{90}\simeq 1$ does not show strong variability which disfavors the internal shock model.}  The principal motivation for evoking internal shocks is related to the observation of variable $\gamma$-ray light curves.  In the framework of internal collisions more than one $\gamma$-ray peak is expected, with a variability timescale  much shorter than the duration of the main activity $\delta t_{\rm var}/t\ll 1$ \citep{1997ApJ...490...92K, 1994ApJ...430L..93R, 2005Sci...309.1833B}.  The properties of several light curves exhibiting one single peak without variability have been explained in the framework of forward/reverse shocks and high-latitude emission \citep[i.e. GRB970508, GRB021211, GRB050406 and others;][]{2003MNRAS.346..905K, 2004MNRAS.354..915M, 2006MNRAS.366..575M, 2004MNRAS.353..647N, 2007ApJ...655..391K, 2009ApJ...698..417P, 2010ApJ...720.1146L}. In the case of the short-lived reverse shock, it can generate a $\gamma$-ray, X-ray or optical peak with $\delta t_{\rm var}/t\simeq 1$  depending on microphysical parameters and the circumburst density  \citep{2007ApJ...655..391K, 2016ApJ...818..190F}.  Therefore,  the emission generated by the reverse shock could in principle describe naturally the variability timescale of  the $\gamma$-ray light curve.\\
\\
\paragraph {2. The value of temporal index $\alpha_\gamma=2.85\pm0.35$  observed during the peak decay phase  is consistent with the high-latitude afterglow emission.}  The most adopted  interpretation to account for the peak decay phase in optical, X-ray and $\gamma$-ray bands, is attributed to delayed photons arriving from high latitudes (curvature effect).  \cite{2000ApJ...541L..51K} showed that the evolution of the observed flux, when it is originated at high latitudes, is $F_{\rm obs}\propto t^{\beta-2}$.  The values of the spectral index, $\beta$, correspond to the low -$\frac12$ ($\frac{1-p}{2}$) and high-energy -$\frac{p}{2}$ (-$\frac{p}{2}$) photon indexes of the synchrotron self-Compton (SSC)/synchrotron spectrum in the fast (slow) cooling regime.  Taking into consideration the typical values of the spectral power index for external shocks, $2.2\leq p \leq 2.6$  \cite[e.g., see;][]{2015PhR...561....1K}, the high-latitude afterglow flux is expected to evolve as $F_\nu\propto t^{-\alpha}$ with  $2.5\leq\alpha \leq 3.2$ which is in accordance with the value obtained of peak decay index $\alpha_\gamma=2.85\pm0.35$. Similar results have been found in a large determined group of GRBs when the peak has been modelled through SSC/synchrotron reverse-shock emission at high latitudes \citep[i.e, see][]{2006ApJ...647.1213O, 2007ApJ...655..391K, 2016ApJ...831...22F, 2006ApJ...642..354Z, 2005ApJ...628..315Z, 2003ApJ...595..950Z, 2003ApJ...597..455K, 2017ApJ...848...15F, 2017ApJ...848...94F}.   On the other hand,  based on observations the typical values of decay index phase  associated to internal shocks are $5\lesssim\alpha \lesssim 7$ \citep{2006ApJ...642..354Z, 2006RPPh...69.2259M, 2015PhR...561....1K, 2005MNRAS.364L..42F} which decays faster than that observed in this burst.\\
\\
\paragraph {3.  The evolution of peak energy at hundreds of keVs during the decay phase $q=0.97\pm 0.35$ \citep{2018arXiv180207328V} is consistent with the SSC energy break of the reverse shock from high latitudes.}  Several authors have studied the spectral evolution of distinct pulses  during  $\gamma$-ray prompt emission. By analyzing the peak decay phase during the prompt emission, some bursts have provided evidence of the synchrotron emission in the fast- and slow-cooling regime from external shocks \citep{1999ApJ...524L..47G, 2003ApJ...595..950Z,2007MNRAS.379..331P,  2017ApJ...848...94F, 2017ApJ...848...15F}.\\   
When SSC/synchrotron spectral breaks are observed coming from high latitudes/off-axis ($\epsilon^{\rm i, off}$ for syn or SSC),   \cite{2017arXiv171005905I} proposed that these spectral breaks must be re-scaled as $\epsilon^{\rm i, off}\propto  \Gamma^{-2}  \epsilon^{\rm i, on}$ with $\Gamma$ the bulk Lorentz factor and $ \epsilon^{\rm syn, on}$ the energy break viewed in an on-axis outflow.  Following \cite{2000ApJ...545..807K}, \cite{2016ApJ...831...22F} and \cite{1998ApJ...497L..17S},   we re-scale the quantities associated to the internal and external shocks.\\
i) Internal shocks: Given the bulk Lorentz factor in the coasting phase  $\Gamma \propto t^0$ and the magnetic field in the freezing regime $B \propto t^0$,  the synchrotron emission evolves as  $\epsilon^{\rm syn, off}_{\rm pk}\propto \Gamma^{-2}  \left(\Gamma\, \gamma_e^2\, B\right)\propto B^{-3}\,\Gamma^{-3}\,t^{-2}\propto t^{-2}$ \citep{2007Ap&SS.309..157D}.  We consider a  ``typical" electron Lorentz factor as one that has the average  of the electron distribution $\langle \gamma_e\rangle=\frac{U_e}{m_e\,N_e}$ \citep{1999PhR...314..575P}.  These quantities can be calculated using two different ways:  1. The energy density given to accelerate electrons is  ${\small {\rm U_e}=\epsilon_e\,U=\epsilon_e\,\gamma_{\rm sh}\,{\rm N_p\,m_p}}$ and the electron number density can be estimated as  ${\rm N_e}\simeq {\rm N_p}$  \citep{1999PhR...314..575P}. In this case, the average electron Lorentz factor  becomes {\small $\langle \gamma_e\rangle=\frac{m_p}{m_e}\epsilon_e \gamma_{\rm sh}$}. 2. The energy density given to accelerate electrons is ${\rm U_e}=\frac{m_e\,A_e}{(p-2)}\gamma^{-p+2}_{\rm e,mi}$ and the electron number density can be estimated as  ${\rm N_e}=\frac{m_e\,A_e}{(p-1)}\gamma^{-p+1}_{\rm e,mi}$ for $p>2$ and $\gamma_{\rm e,mi}\ll\gamma_{\rm e,ma}$.  Therefore,  in this case $\langle \gamma_e\rangle=\frac{p-1}{p-2}\,\gamma_{\rm e, mi}$. Here,  $\gamma_{\rm mi}$ and $\gamma_{\rm ma}$ are the minimum and maximum electron Lorentz factors, respectively, and $\gamma_{\rm sh}$ is the relative Lorentz factor across the internal shock \citep{1999PhR...314..575P}.  Considering both cases and the magnetic field given by $B\simeq \sqrt{\gamma_{\rm sh}}\epsilon_B^{1/2}\Gamma^{-3}\,L^{1/2}_j\, t^{-1}_\nu$, the electrons accelerated and cooled down in internal shocks via synchrotron radiation reach a peak energy at \citep[e.g. see][]{2017ApJ...848...94F}
{\small
\bary
\epsilon^{\rm syn, on}_{\rm pk}&\simeq&
\cases{
0.4\,{\rm MeV}\,\,\varepsilon^2_{e,-0.3} \gamma^2_{\rm sh}\cr
0.6\,{\rm MeV}\,\,\gamma^2_{\rm mi,3}\cr
}\cr
&& \hspace{1.1cm}\times \left(\frac{1.01}{1+z}\right)\sqrt{\gamma_{\rm sh}}\varepsilon^{1/2}_{B,-1}\,\delta t^{-1}_{\rm var,0}  \Gamma^{-2}_{\rm 3} L_{j,49}^{1/2}\,.
\eary
}
The microphysical parameters $\varepsilon_{\rm e}$ and $\varepsilon_{\rm B}$ are the fractions of energy given to accelerate electrons and generate/amplify the magnetic field, respectively, and  $L_{\rm j}$ is the jet luminosity.  Hereafter, the convention $Q_{\rm x}=Q/10^{\rm x}$ in c.g.s. units is adopted.  Including pair formation, an upper limit for the peak energy can be estimated as \citep{2001ApJ...557..399G}
{\small
\be
\epsilon^{\rm syn, on}_{\rm pk}\lesssim 3.3\, {\rm MeV}\,  \left(\frac{1+z}{1.01} \right)^{-1}   \, L^{-1/5}_{j,49} \Gamma^{\frac43}_{\rm 3}\,t^{\frac16}_{\nu,0} \varepsilon^{\frac12}_{B,-1}\,\varepsilon^{\frac43}_{e,-0.3}\,.
\ee
}
The peak energy observed  from high latitudes/off-axis jet has to be rescaled by {\small $\epsilon^{\rm off}_{\rm pk}\simeq b^{-1} \epsilon^{\rm syn, on}_{\rm pk}$} with $b=1+\Gamma^2\Delta\theta^2$ and $\Delta\theta= \theta_{\rm obs} - \theta_j$.  In this case, the observed energy {\small $\epsilon^{\rm syn, off}_{\rm pk}\simeq 10\, {\rm eV}\, \Gamma^{-2}_3\Delta\theta^{-2}_{15^\circ}$}  indicates that it can hardly reach values as high as hundreds of keVs.   Therefore, the standard internal shocks cannot straightforwardly explain the evolution with time of peak energy at hundreds of keVs.\\
\\
ii) Forward shock: Given the evolution of the magnetic field $B\propto t^{-3/8}$,  the bulk Lorentz factor $\Gamma\propto t^{-3/8}$, the  minimum and cooling electron Lorentz factors $\gamma_{\rm m,f}\propto t^{-3/8}$ and $\gamma_{\rm c,f}\propto t^{1/8}$ \citep{1998ApJ...497L..17S}, respectively, the synchrotron spectral breaks evolve as $\epsilon^{\rm syn, off}_{\rm m,f}\propto  B\,\Gamma^{-3}\, \gamma_{\rm em,f} \propto t^{-3/4}$  and $\epsilon^{\rm syn,off}_{\rm c,f}\propto B^{-5}\,\Gamma^{-3}\, t^{-2} \propto t^{1/4}$.   The SSC spectral breaks evolve as $\epsilon^{\rm ssc, off}_{\rm m,f} \propto  \gamma^2_{\rm m,r} \epsilon^{\rm syn, off}_{\rm m,f} \propto t^{-3/2}$  and $\epsilon^{\rm ssc, off}_{\rm c,f} \propto  \gamma^2_{\rm c,f} \epsilon^{\rm syn, off}_{\rm c,f} \propto t^{1/2}$. The subindex ``${\rm f}$" indicates the forward  shock.\\
The synchrotron spectral break $\epsilon^{\rm syn, off}_{\rm m,f}\propto t^{-3/4}$ is the only spectral break that agrees with the peak energy evolution. The synchrotron spectral break is given by 
{\small
\bary\label{synfor_b}\nonumber
\epsilon^{\rm syn, off}_{\rm m,f}&\simeq& 0.3\, \,{\rm keV} \left(\frac{1+z}{1.01}\right)^{\frac12}\,\varepsilon_{\rm e,-0.3}^2\,\varepsilon^{\frac12}_{B,f,-1}\,E^{\frac12}_{51}\, \Gamma^{-2}_{2.5}\,\Delta\theta^{-2}_{10^\circ}\cr
&&\hspace{5.6cm}\times \,t^{-\frac32}_{0}\,,\cr
\eary
}
where  $E=E_{\rm \gamma,iso}/\eta$ is the isotropic equivalent kinetic energy with $\eta$ the kinetic efficiency.   It shows that although synchrotron spectral break  ($\epsilon^{\rm syn, off}_{\rm m,f}$) agrees with the peak energy evolution, it  cannot reach the values of energies at hundreds of keVs. \\
\\
iii) Reverse shock in the thick- (thin-) shell case:   Given the evolution of the magnetic field  $B\propto t^{-1/4} (t^0)$, the bulk Lorentz factor  $\Gamma\propto t^{-1/4}(t^0)$, the  minimum and cooling electron Lorentz factors $\gamma_{\rm m,r}\propto t^{1/4} (t^3)$ and $\gamma_{\rm c,r}\propto t^{-1/4} (t^{-1})$ \citep{2000ApJ...545..807K}, respectively, before the crossing time,  the synchrotron spectral breaks evolve as $\epsilon^{\rm syn,off}_{\rm m,r}\propto t^{1/2}(t^{6})$  and $\epsilon^{\rm syn,off}_{\rm c,r}\propto  t^{-1/2}(t^{-2})$.  The SSC spectral breaks evolve as $\epsilon^{\rm ssc, off}_{\rm m,r} \propto  t(t^{12})$  and $\epsilon^{\rm ssc, off}_{\rm c,r} \propto  t^{-1}(t^{-4})$.    Taking into account that the quantities after the crossing time vary as $B\propto t^{-13/24}(t^{-4/7})$,  $\Gamma\propto t^{-7/16}(t^{-2/5})$, $\gamma_{\rm m,r}\propto t^{-13/48}(t^{-2/7})$ and $\gamma_{\rm c,r}\propto t^{25/48} (t^{19/35})$,   the synchrotron spectral breaks evolve as $\epsilon^{\rm syn,off}_{\rm m,r} \propto t^{-0.65}(t^{-0.74})$  and $\epsilon^{\rm syn,off}_{\rm m,r}\propto t^{0.94}(t^{0.91})$.  The SSC spectral breaks evolve as $\epsilon^{\rm ssc, off}_{\rm m,r} \propto t^{-1.18}(t^{-1.31})$  and $\epsilon^{\rm ssc, off}_{\rm c,r} \propto  t^{1.98}(t^{1.99})$.  The subindex ``${\rm r}$" indicates the reverse shock. \\
\\
The SSC and synchrotron spectral breaks that agree with  the peak energy evolution are $\epsilon^{\rm syn, off}_{\rm m,r} \propto t^{-0.65}(t^{-0.74})$ and  $\epsilon^{\rm ssc, off}_{\rm m,r} \propto t^{-1.18}(t^{-1.31})$, respectively. The SSC and synchrotron spectral breaks are given by \citep{2000ApJ...545..807K, 2016ApJ...831...22F}
{\small 
\bary\label{synrev_b}\nonumber
\epsilon^{\rm syn, on}_{\rm m,r}&\simeq&   16.2\,{\rm eV}  \,\left(\frac{1+z}{1.01}\right)^{-1}\,\varepsilon_{\rm e, -0.3}^{2}\,\varepsilon_{\rm B,r,-1}^{\frac12}\,\Gamma^{2}_{2.8}\,{\rm n}^{\frac12}_{0}, \cr
\epsilon^{\rm ssc, on}_{\rm m,r}&\simeq&13.4\,{\rm MeV} \, \left(\frac{1+z}{1.01}\right)^{-\frac74}\,\varepsilon_{\rm e, -0.3}^{4}\,\varepsilon_{\rm B,r, -1}^{\frac12}\,\Gamma^{4}_{2.8}\,{\rm n}^{\frac34}_0\,E^{-\frac14}_{51}\cr
&& \hspace{5.5cm} \times\, t_{\rm cr, 0}^{\frac34} \,,\cr
\eary
}
which have to be rescaled again by $(1+\Gamma^2 \Delta\theta^2)^{-1}$  \citep{2017arXiv171005905I}. The term  t$_{\rm cr}$ is  the shock crossing time. The parameter ${\rm n}$ corresponds to the circumburst density.  For this analysis, the bulk Lorentz factor corresponds to that one associated with the reverse shock.    This value can be estimated taking into consideration the four-region structure during the shock: (1) the unshocked ISM with density $n_1$, (2) the shocked ISM, (3) the shocked shell material and (4) the unshocked shell material with density $n_4$ and  the equations governing the shocks with the jump conditions  $\frac{n_4}{n_1}\,\simeq\frac{(\gamma_3-1)(4\gamma_3+3)}{(\gamma_{34}-1)(4\gamma_{34}+3)}$ and  $\gamma_{34}\simeq \frac12\left(\frac{\gamma_4}{\gamma_3}+ \frac{\gamma_3}{\gamma_4} \right)$ \citep{1976PhFl...19.1130B, 1995ApJ...455L.143S}, with $\gamma_{34}$ the relative Lorentz factor between the upstream and downstream region, $\gamma_3\equiv \Gamma_r$ and $\gamma_4\equiv \Gamma$ the reverse and initial Lorentz factors, respectively \citep{2000ApJ...545..807K}.  For the relativistic case,  i.e., $\gamma_{34}\gg 1$, the  bulk Lorentz factor of the reverse shock is
\be 
\Gamma_r\simeq  \sqrt{\frac{\Gamma}{2}} \left(\frac{n_4}{n_1} \right)^{1/4}\,.
\ee
For typical values of the initial bulk Lorentz factor $300 \leq \Gamma \leq 600$ and densities of unshocked ISM and shell $n_4\simeq n_1$, the bulk Lorentz factor becomes $12.3\leq \Gamma_r\leq 17.3$. For the case of $\Gamma_r=15$, SSC and synchrotron spectral breaks have to be rescaled by $\approx 7\times 10^{-2} \Gamma^{-2}_{\rm r, 1.2}\,\Delta\theta^{-2}_{15^\circ}$. Therefore, the characteristic break of SSC reverse-shock emission agrees with   the evolution of peak energy at hundreds of keVs during the decay phase.\\
\vspace{0.8cm}
\paragraph {4. The  hard-to-soft spectral-index evolution (from -1.749$\pm$0.434  to -2.150$\pm$0.472) seems to be consistent with SSC/synchrotron spectrum originated in external shocks.}  Ultra-relativistic electrons confined in a magnetic field are cooled down by synchrotron and SSC radiation. The high and low spectral indexes in the fast(slow)-cooling regime are -$\frac12$(-$\frac{p-1}{2}$) and -$\frac{p}{2}$(-$\frac{p}{2}$), respectively. Given the typical values of the spectral power index for external shocks, $2.2\leq p \leq 2.6$  \cite[e.g., see;][]{2015PhR...561....1K}, the SSC/synchrotron spectrum $\nu F_\nu\propto \nu^{-(\beta+1)}$ with $1.5(1.6)\leq\beta+1\leq 2.3(2.3)$ agrees with the spectral-index evolution for fast(slow)-cooling regime.\\
\\
The previous analysis, performed on the  temporal and spectral features of the $\gamma$-ray light curve (see Figure \ref{fig1:gbm} and Tables \ref{table1:gbm_analysis} and \ref{table2: fit_gbm}), illustrates that:  i) The characteristic break of SSC reverse shock agrees with  the evolution of peak energy at hundreds of keVs during the decay phase, while synchrotron emission from internal and forward shocks cannot explain this evolution,  ii) The reverse-shock emission can reproduce, in a more natural way, the observed variability timescale than internal-shock emission and  iii) The temporal and spectral indexes of synchrotron/SSC emission, originated from external shocks, are consistent with the spectral-index evolution and the high-latitude afterglow model. Therefore, we argue that the SSC reverse-shock emission in the fast-cooling regime reproduce  the  temporal and spectral features of the $\gamma$-ray light curve. In the following subsection the SSC spectrum in the fast cooling regime is used to describe the the $\gamma$-ray flux.\\
\subsubsection{Theoretical model}
\begin{figure}
{ \centering
\resizebox*{1\textwidth}{0.7\textheight}
{\includegraphics[angle=-90]{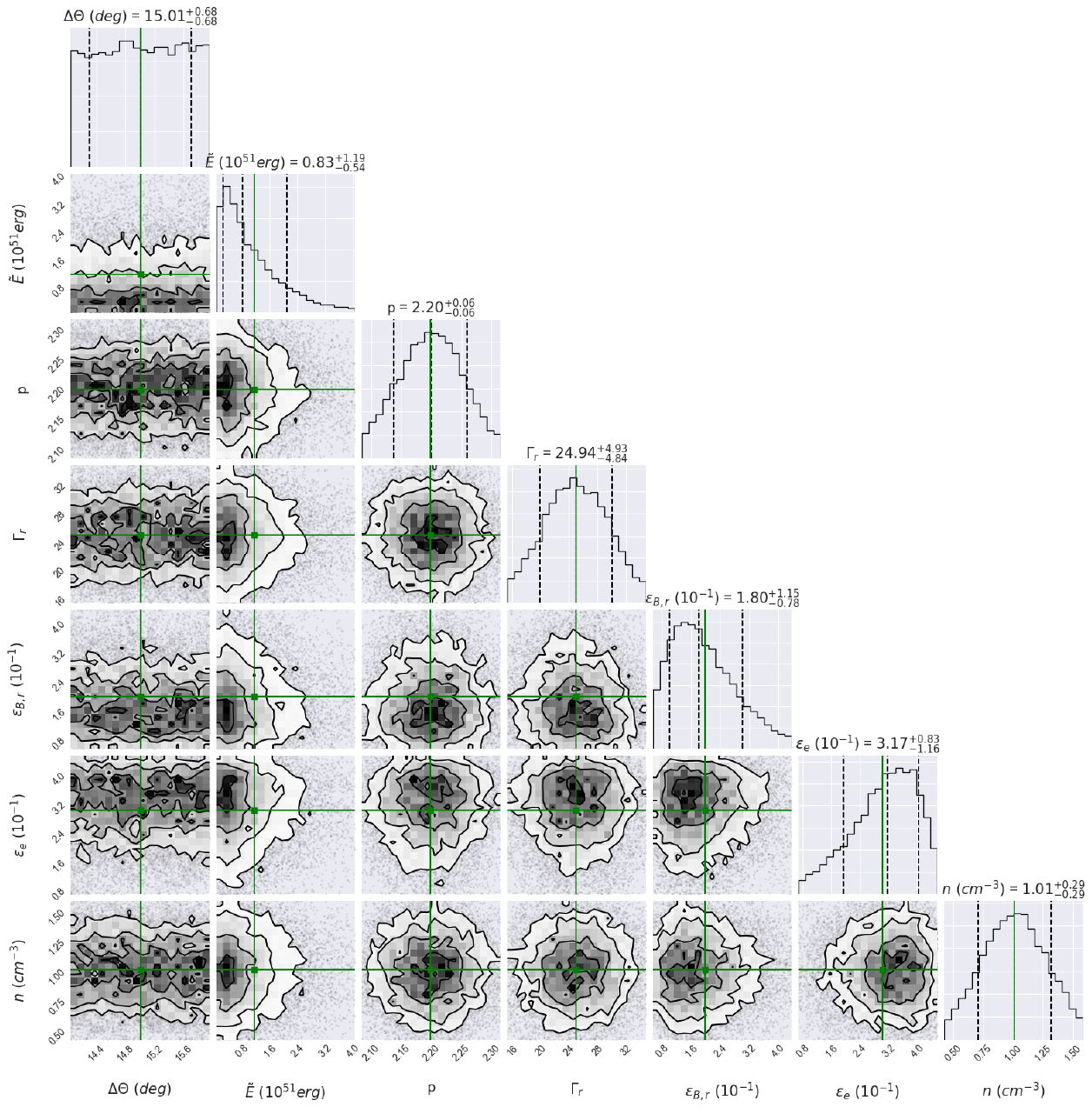}}
}
\caption{Corner plot compares the results obtained from the MCMC simulation for each parameter.   Fit result for GBM $\gamma$-ray data using a SSC reverse shock model in a homogeneous density as described in section 2.1. Labels above the 1-D kernel plots indicate the median, 0.15 and 0.85 quantiles of each parameter.   The best-fit value is shown in green color.}
\label{fig2:param_early}
\end{figure}

 The SSC spectral breaks and  fluxes are determined by the spectral break evolution between  forward and reverse shocks \citep{2005ApJ...628..315Z, 2015ApJ...804..105F, 2016ApJ...818..190F}.  The SSC spectrum in the fast cooling regime is given by \citep{2017arXiv171005905I}
{\small
\be
F^{\rm ssc, on}_{\nu,r}= F^{\rm ssc, on}_{\rm max,r}\left(\frac{\epsilon_\gamma}{\epsilon^{\rm ssc, on}_{\rm c,r}} \right)^{-\frac{1}{2}}\,\, {\rm for}\,\, \epsilon^{\rm ssc, on}_{\rm c,r}<\epsilon_\gamma<\epsilon^{\rm ssc, on}_{\rm m,r}\,,
\ee
}
and once the characteristic break energy passes through the $\gamma$-ray band  $\epsilon_\gamma\approx$ 100 keV at $\sim t_{\rm cr} \epsilon^{\rm ssc, on}_{\rm m,r}/\epsilon_\gamma$, the SSC flux begins evolving in the following power-law segment of the light curve  {\small $F^{\rm ssc, on}_{\rm max,r}\,\left(\frac{\epsilon^{\rm ssc, on}_{\rm m,r}}{\epsilon^{\rm ssc, on}_{\rm c,r}} \right)^{-\frac{1}{2}}\,\left(\frac{\epsilon_\gamma}{\epsilon^{\rm ssc, on}_{\rm m,r}}\right)^{-\frac{p}{2}}$} for $ \epsilon^{\rm syn, on}_{\rm c,r}< \epsilon^{\rm syn, on}_{\rm m, r}<\epsilon_\gamma$.   The SSC energy breaks and the maximum fluxes when reverse shock evolves in the thick shell are given explicitly in \cite{2012ApJ...751...33F}.  These quantities viewed off-axis must be corrected by 
%
%
{\small
\bary
\epsilon^{\rm ssc,off}_{\rm m/c,r}= b^{-1}\epsilon^{\rm ssc,on}_{\rm m/c,r},\,\hspace{0.5cm} {\rm and}\,\, \hspace{0.5cm} F^{\rm ssc, off}_{\rm max,r}= b^{-3} F^{\rm ssc, on}_{\rm max,r}\,.
\eary
}
To find the best-fit values that reproduce the data with our off-axis model,  we perform the Bayesian statistical method of Markov-Chain Monte Carlo (MCMC) simulations. Our model is fully described by a set of seven parameters, $\Xi_{\rm rev}$ = \{n, $\tilde{E}$, $\Delta\theta$, $p$, $\Gamma_r$, $\varepsilon_{B,r}$, $\varepsilon_e \}$, with an extra parameter of $\sigma$ for the likelihood of the MCMC.   We generate samples of the posterior for our off-axis model using the No-U-Turn Sampler (NUTS) from the PyMC3 python distribution \citep{peerj-cs.55}. In order to fit data,  we run the model for fluxes with a total of 14000 samples and 3000 tuning steps, which are to be discarded after tuning.   The priors are assigned independently, with a mixture of different continuous probability distributions functions and standard deviations. The parameters $p$, $\Gamma_r$ and $n$ are given normal distributions, $\Delta\theta$ a uniform distribution and $\varepsilon_{B,r}$, $\varepsilon_{e}$ and $\tilde{E}$ are given modified normal distributions.   Output is given by means of a Corner Plot \citep{2016JOSS....1...24F} on which the diagonal is a one-dimensional kernel plot of the posterior probability distribution function and the off-diagonal are the bi-dimensional kernel plots. Hereafter, the values of cosmological parameters $H_0=$ 71 km s$^{-1}$ Mpc$^{-1}$, $\Omega_m=0.27$, $\Omega_\lambda=0.73$  are adopted \citep{2003ApJS..148..175S}.\\ 
Figure \ref{fig2:param_early} shows the corner plots  obtained from the MCMC simulation for each parameter  using  SSC reverse shock model.   The best-fit values in this figure are shown in green color.  The median of the posterior distributions, alongside with the symmetrical 35\% quantiles, are  reported in Table \ref{table3:param_early}. 
\vspace{1cm}
\subsection{Modeling the non-thermal X-ray, optical and radio data}

\subsubsection{Light curve analysis and description}
Several X-ray observations were carried out during the following 8 days after the merger providing constraining limits \citep[i.e. see][]{2017ApJ...848L..20M}. On the ninth day, the Chandra X-ray observatory reported a faint X-ray flux from the direction of the binary NS merger \citep{troja2017a}. From the 108th to 256th day post GW trigger, Chandra and XMM-Newton observatories reported  detections \citep{2017ATel11037....1M, 2018ATel11242....1H}.  The Hubble Space Telescope (HST) observed optical non-thermal fluxes with magnitudes $26.44\pm 0.14$ mag \citep{2018arXiv180102669L} and $26.90\pm 0.25$ mag \citep{2018arXiv180103531M} at $\sim$ 110 and 137 days, respectively, after the merger. On 2018 March 23, HST provides an upper limit of $>0.070\,{\rm \mu Jy}$ \citep{2018arXiv180502870A}.    Since the sixteenth day post-trigger and for more than seven months, Very Large Array (VLA) has reported a faint radio flux at 3 and 6 GHz \citep{2017Natur.547..425T, 2017arXiv171005435H, 2017arXiv171111573M, 2017ApJ...848L..21A}.\\
In order to describe, firstly, the X-ray, optical and radio light curves during the increasing phase,  we consider these  light curves up to 145 $\pm$ 20 days and the broadband SED in three separate periods: at 15 $\pm$ 2,  110 $\pm$ 5 and 145 $\pm$ 20 days.
The X-ray and radio (6 and 3 GHz) light curves up to 145 $\pm$ 20 days were adjusted to simple power laws  $F_\nu\propto t^{-\alpha_i}$ (for i=X,  6GHz and 3GHz), and the broadband SED at 15 $\pm$ 2,  110 $\pm$ 5 and 145 $\pm$ 20 days were fitted with  $F_\nu\propto \nu^{-\beta_t}$. The best-fit values of temporal and spectral indexes obtained with the $\chi^2$ test implemented in the ROOT software package are reported in Table \ref{table4:fit_root}.  Given the best-fit values obtained up to 145 $\pm$ 20 days, the multiwavelength fluxes can be described as  $\propto t^{0.76\pm 0.18}\,\nu^{-0.58\pm0.15}$ for X-ray, optical and radio data. \\
\\
Afterglow emission is generated when the relativistic jet encounters the homogeneous medium and sweeps up enough circumburst material.  The  synchrotron forward-shock model are the most favorable one to describe the late time multiwavelength observations.   Taking into account the closure relations of the standard synchrotron forward-shock model,  the X-ray, optical and radio (6 and 3 GHz) fluxes are evolving in the slow-cooling regime  corresponding to the power-law segment  \citep{1998ApJ...497L..17S}
 {\small
\be\label{in_flux}
F^{\rm syn,on}_{\nu,f}= F^{\rm syn,on}_{\rm max,f}\left(\frac{\epsilon_\gamma}{\epsilon^{\rm syn,on}_{\rm m,r}} \right)^{-\frac{p-1}{2}}\,\, {\rm for}\,\, \epsilon^{\rm syn}_{\rm m,r}<\epsilon_\gamma<\epsilon^{\rm syn, on}_{\rm c,r}\,,
\ee
}
with $p=2\beta+1\approx2.2$  and

\be\label{standardAfter}
 \epsilon^{\rm syn,on}_{\rm m, f}\propto t^0\,\Gamma^4\,,\hspace{0.4cm}   \epsilon^{\rm syn,on}_{\rm c, f}\propto \,t^{-2}\,\Gamma^{-4}\hspace{0.3cm}   {\rm and} \hspace{0.3cm}  F^{\rm syn,on}_{\rm max, f} \propto E\,t^0 \Gamma^0\,.
 \ee
Considering the evolution of the bulk Lorentz factor, {\small $\Gamma\propto \,t^{-\frac38}$}, the flux varies as $F^{\rm syn, on}_{\rm \nu,f}\propto t^{-\frac{3(p-1)}{4}}$. Given the observed temporal index reported in Table \ref{table4:fit_root}, the value of electron distribution would be $p\approx-0.07$ which is inconsistent with the value obtained from the broadband SED ($p\approx 2.2$).  While the evolution of the synchrotron flux as function of the energy is well-described, the evolution of it with time fails.  This inconsistency is due to the evolution of the bulk Lorentz factor.\\
Given that the standard synchrotron afterglow model cannot account for the X-ray, optical and radio light curves of GRB 170817A,  we consider the synchrotron forward-shock model to be off-axis when the matter in the outflow is parametrized  through a power law velocity distribution.
\vspace{0.3cm}
\subsubsection{Theoretical model}
\begin{figure}
{ \centering
\resizebox*{\textwidth}{0.7\textheight}
{\includegraphics[angle=-90]{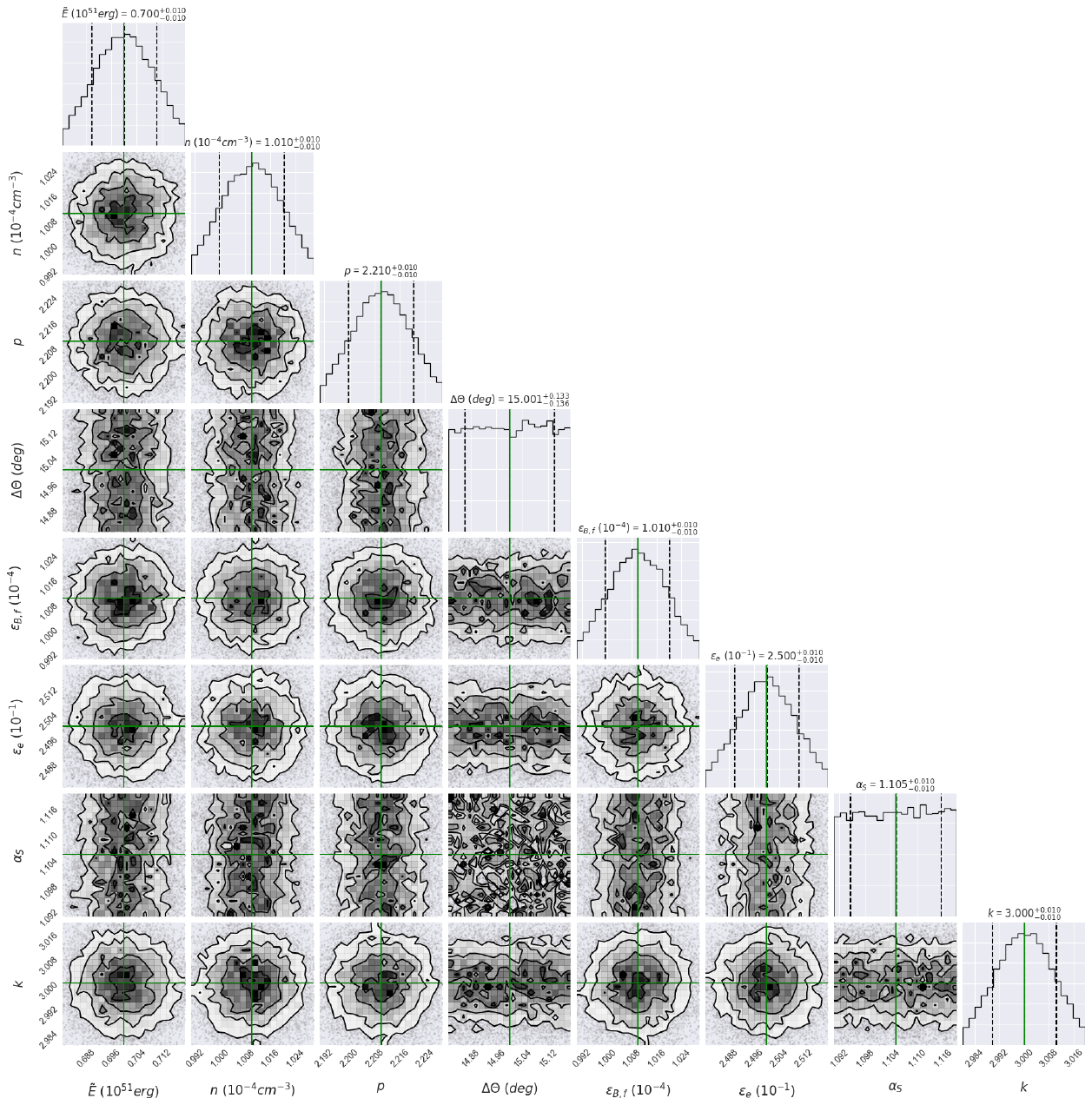}}
}
\caption{Same as Fig. \ref{fig2:param_early}, but using a synchrotron forward-shock model to fit the radio (3 GHz) data when the jet producing the afterglow emission is not aligned with the observer's line of sight  and the matter in the outflow is  parametrized through  a power law velocity distribution (model described in section 2.2).}
\label{fig3:param_late_3R}
\end{figure}

\begin{figure}
{ \centering
\resizebox*{\textwidth}{0.7\textheight}
{\includegraphics[angle=-90]{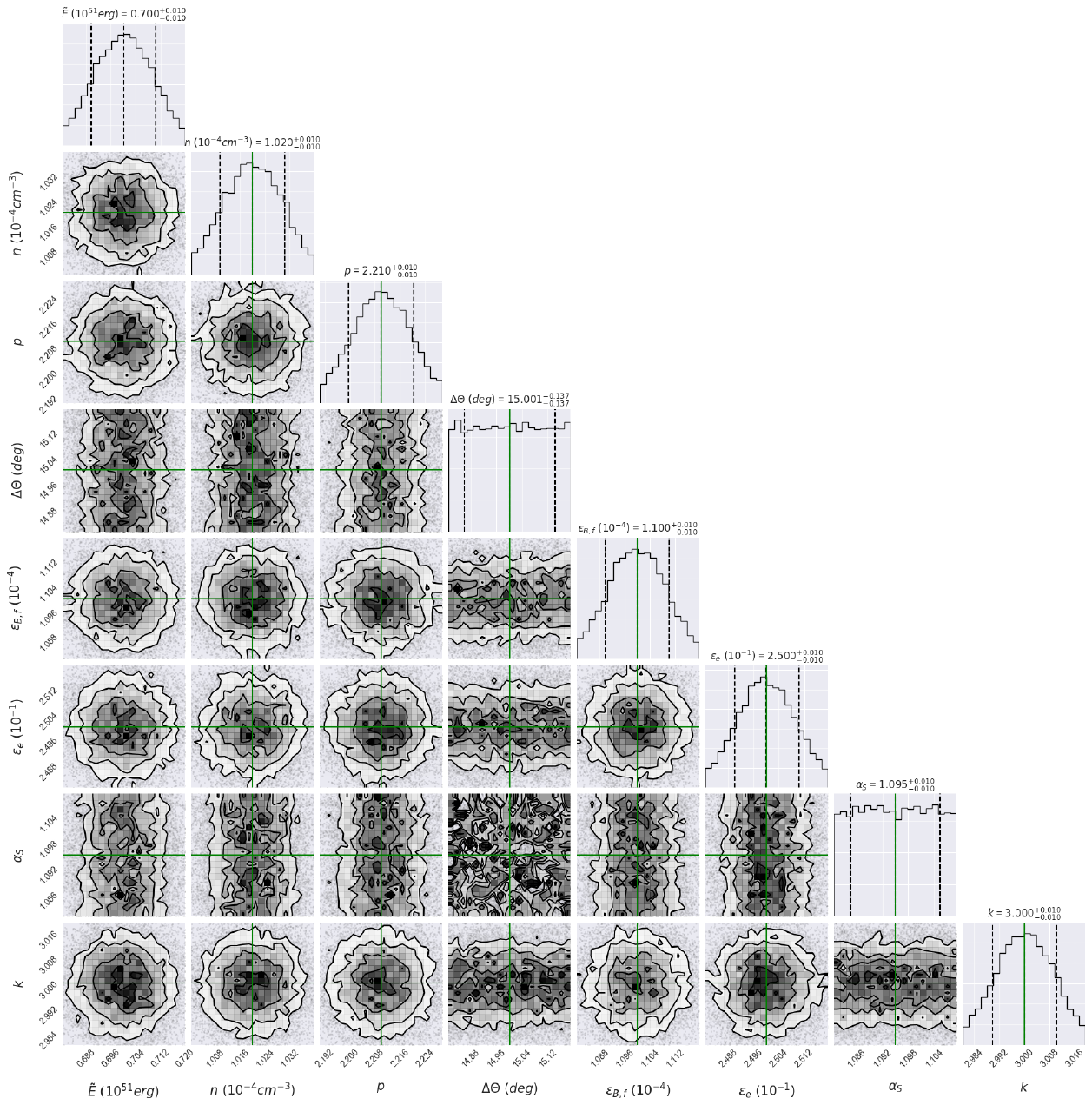}}
}
\caption{Same as Fig. \ref{fig2:param_early}, but using a synchrotron forward-shock model to fit the radio (6 GHz) data when the jet producing the afterglow emission is not aligned with the observer's line of sight  and the matter in the outflow is  parametrized through  a power law velocity distribution (model described in section 2.2).}
\label{fig3:param_late_6R}
\end{figure}

\begin{figure}
{ \centering
\resizebox*{\textwidth}{0.7\textheight}
{\includegraphics[angle=-90]{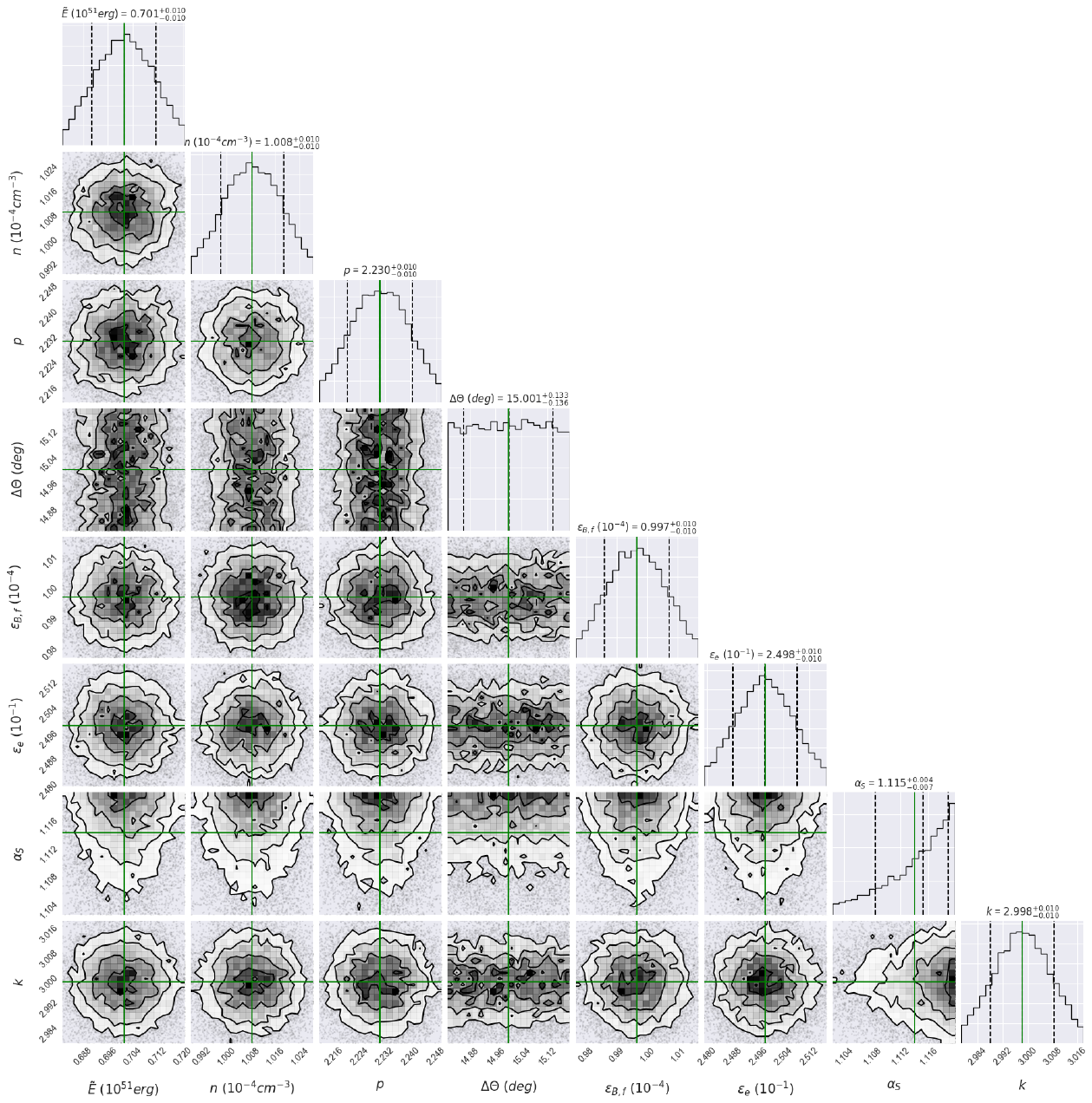}}
}
\caption{Same as Fig. \ref{fig2:param_early}, but using a synchrotron forward-shock model to fit the X-ray data when the jet producing the afterglow emission is not aligned with the observer's line of sight  and the matter in the outflow is  parametrized through  a power law velocity distribution (model described in section 2.2).}
\label{fig3:param_late_X}
\end{figure}

We consider that the jet concentrated within an opening angle $\theta_j$  ``top-hat jet" producing the afterglow emission is not aligned with the observer's line of sight and the ejecta  has an equivalent kinetic energy  parametrized by a power law distribution as  {\rm $ \tilde{E}\,\left(\beta\Gamma\right)^{-\alpha_s}$} where $\tilde{E}$ is the fiducial energy, $\alpha_s=1.1$ for $\beta\Gamma\gg 1$ and $\alpha_s=5.2$ for $\beta\Gamma\ll 1$ for the adiabatic case    \citep{2001ApJ...551..946T, 2000ApJ...535L..33S, 2015MNRAS.448..417B, 2015MNRAS.450.1430H, 2013ApJ...778L..16H, 2014MNRAS.437L...6K,  2018arXiv180302978F}.  Taking into account the relativistic regime ($\beta \Gamma\gg1$),  we propose that the corresponding equivalent kinetic energy for $\theta_{\rm obs}\gtrsim 2\theta_j$ is given by
\bary\label{Eoff}
E_{\rm k}&=& b^{-3}\,\tilde{E}\,\Gamma^{-\alpha_s}  \cr
&\simeq& \Delta \theta^{-6} \Gamma^{-\delta} \tilde{E}\,,
\eary
%
%
%
%
%
%
for $\Gamma^2 \Delta \theta^2\gg 1$ with $\Delta\theta= \theta_{\rm obs} - \theta_j$   and $\delta=\alpha_s+6$.\\
Considering the adiabatic evolution of the forward shock \citep{1976PhFl...19.1130B, 1997ApJ...489L..37S}, the fiducial energy is given by $\tilde{E}=16/17\pi \Delta \theta^6 \Gamma^{\delta+2}R^3n\,m_p$ \citep{1976PhFl...19.1130B, 1997ApJ...489L..37S} with  $m_p$ the proton mass and  $R$ the deceleration radius.  In this case, the bulk Lorentz  factor evolves as
{\small
\be\label{newLorentz}
\Gamma = 7.8\,\left(\frac{1+z}{1.01} \right)^\frac{3}{\delta+8}\,n^{-\frac{1}{\delta+8}}_{-4}\, \,\tilde{E}^{\frac{1}{\delta+8}}_{51}\,\Delta \theta^{-\frac{6}{\delta+8}}_{20^\circ}\, t^{-\frac{3}{\delta+8}}_{1\,{\rm d}}\,.
\ee
}
Replacing eqs. (\ref{newLorentz})  and (\ref{Eoff}) in (\ref{standardAfter}), the synchrotron spectral breaks and the maximum flux are
{\small
\bary\label{energies_break}
\epsilon^{\rm syn}_{\rm m,f}&\simeq& 7.7\times 10^{-4}\,{\rm GHz}\,\, \left(\frac{1+z}{1.01} \right)^{\frac{4-\delta}{\delta+8}}\varepsilon^2_{e,-1}\,\varepsilon_{B,-4}^{\frac12}\,n_{-4}^{\frac{\delta}{2(\delta+8)}}\, \cr
&&\hspace{3.8cm} \times\,\tilde{E}^{\frac{4}{\delta+8}}_{51}\,\Delta \theta^{-\frac{24}{\delta+8}}_{20^\circ}\, t_{100\,{\rm d}}^{-\frac{12}{\delta+8}} \hspace{1cm}\cr
\epsilon^{\rm syn}_{\rm c, f}&\simeq&  5.2\,{\rm keV}  \left(\frac{1+z}{1.01} \right)^{\frac{\delta-4}{\delta+8}} (1+x)^{-2}\, \varepsilon_{B,-4}^{-\frac32}\,n_{-4}^{-\frac{16+3\delta}{2(\delta+8)}}\cr
&&\hspace{3.5cm} \times\,\tilde{E}^{-\frac{4}{\delta+8}}_{51}\, \Delta \theta^{\frac{24}{\delta+8}}_{20^\circ}\, t_{100\,{\rm d}}^{-\frac{2\delta+4}{\delta+8}}\cr
F^{\rm syn}_{\rm max, f} &\simeq& 1.4\,{\rm mJy}\,\, \left(\frac{1+z}{1.01} \right)^{\frac{8-2\delta}{\delta+8}}\,\varepsilon_{B,-4}^{\frac12}\,n_{-4}^{\frac{3\delta+8}{2(\delta+8)}}\, D^{-2}_{26.1}  \,\tilde{E}^{\frac{8}{\delta+8}}_{51}\cr
&&\hspace{4.4cm} \times\, \Delta \theta^{-\frac{48}{\delta+8}}_{20^\circ}\,t_{100\,{\rm d}}^{\frac{3\delta}{\delta+8}}\,.\cr
\eary
}
Given the new evolution of the synchrotron emission, from eqs. (\ref{energies_break}) and (\ref{in_flux}), the  power-law segment of the synchrotron spectrum in the slow-cooling regime becomes
{\small
\bary
F_{\rm \nu, inc}&\simeq&
F_{\rm \nu, i}\,\, t^{\frac{3\delta-6(p-1)}{\delta+8}}_{100\,{\rm d}}\, \epsilon_\gamma^{-\frac{p-1}{2}}\, A_{\rm \nu, inc}\,,
\eary
}
where
{\small
\bary
A_{\rm \nu, inc}&=&   \left(\frac{1+z}{1.01}\right)^{-\frac{7\delta+12+p\delta+4p}{2(\delta+8)}}\,\varepsilon^{p-1}_{e,-1}\,\varepsilon^{\frac{p+1}{4}}_{B,-4}\,n_{-4}^{\frac{16+ \delta(p+5)}{4(\delta+8)}}\,D^{-2}_{26.1}\,\cr
&&\hspace{3cm}\times\,\tilde{E}^{\frac{6+2p}{\delta+8}}_{51}\,\Delta \theta^{-\frac{12(p+3)}{\delta+8}}_{20^\circ}\,\,,
\eary
}
and {\small $F_{\rm \nu, i}=\{9.8\times \,10^{-3}$, $6.5\times \,10^{-3}$, $1.1\times \,10^{-5}$,  $1.8\times \,10^{-7}\}$ {\rm mJy}} for {\small $\epsilon_{\rm \gamma}= \{3\, {\rm GHz}$,  $6\, {\rm GHz}$,  $1\, {\rm eV}$,  $1\, {\rm keV}\}$}, respectively.
For this case, the flux varies as $F_{\nu}\propto\,t^{\frac{3\delta-6(p-1)}{\delta+8}}\,\nu^{-\frac{p-1}{2}}$, which for $\alpha_s\approx1.1$ and $p\approx2.2$  it evolves as  found after fitting the SED at 15 $\pm$ 2,  110 $\pm$ 5 and 145 $\pm$ 20 days and reported in Table \ref{table4:fit_root}.  It is worth noting that for $\delta=0$, the flux $F_{\rm \nu, dec}\propto t^{-\frac{3(p-1)}{4}}$ derived in \cite{1998ApJ...497L..17S} is recovered.\\
Since the radiation beaming cone  broadens increasingly, it reaches our line of sight later \citep[$\Gamma\sim\Delta \theta^{-1}$; ][]{2000ApJ...537..785D, 2002ApJ...570L..61G, 1999A&AS..138..491R, 2017arXiv171006421G, 1999ApJ...519L..17S}.  Once the flux reaches our field of view the synchrotron spectral breaks and the maximum flux become
{\small
\bary\label{energies_break}
\epsilon^{\rm syn}_{\rm m,f}&\simeq& 2.9\times 10^{-4}\,{\rm GHz}\,\, \left(\frac{1+z}{1.01} \right)^{\frac{6-\alpha_s}{\alpha_s+6}}\varepsilon^2_{e,-1}\,\varepsilon_{B,-4}^{\frac12}\,n_{-4}^{\frac{\alpha_s-2}{2(\alpha_s+6)}}\, \cr
&&\hspace{4.6cm} \times\,\tilde{E}^{\frac{4}{\alpha_s+6}}_{51}\, t_{200\,{\rm d}}^{-\frac{12}{\alpha_s+6}} \hspace{1cm}\cr
\epsilon^{\rm syn}_{\rm c, f}&\simeq&  3.4\,{\rm keV}  \left(\frac{1+z}{1.01} \right)^{\frac{\alpha_s-6}{\alpha_s+6}} (1+x)^{-2}\, \varepsilon_{B,-4}^{-\frac32}\,n_{-4}^{-\frac{3\alpha_s+10}{2(\alpha_s+6)}}\cr
&&\hspace{4.5cm} \times\,\tilde{E}^{-\frac{4}{\alpha_s+6}}_{51}\, t_{200\,{\rm d}}^{-\frac{2\alpha_s}{\alpha_s+6}}\cr
F^{\rm syn}_{\rm max, f} &\simeq& 1.1\,{\rm mJy}\,\, \left(\frac{1+z}{1.01} \right)^{-\frac{4\alpha_s}{\alpha_s+6}}\,\varepsilon_{B,-4}^{\frac12}\,n_{-4}^{\frac{3\alpha_s+2}{2(\alpha_s+6)}}\, D^{-2}_{26.1}  \,\tilde{E}^{\frac{8}{\alpha_s+6}}_{51}\cr
&&\hspace{4.6cm} \times\,t_{200\,{\rm d}}^{-\frac{3(2-\alpha_s)}{\alpha_s+6}}\,.\cr
\eary
}
Since  $\epsilon^{\rm syn}_{\rm m,f}\leq \epsilon_\gamma   \leq \epsilon^{\rm syn}_{\rm c, f}$, the flux lies in the same power-law segment. It begins decreasing  as  
{\small
\bary
F_{\rm \nu, dec}&\simeq&
F_{\rm nu, j}\,\,t^{-\frac{3(\alpha_s-2p)}{\alpha_s+6}}_{200\,{\rm d}}\, \epsilon_\gamma^{-\frac{p-1}{2}}\, A_{\rm \nu, dec}\,,
\eary
}
where
{\small
\bary
A_{\rm \nu, dec}&=&\left(\frac{1+z}{1.01}\right)^{\frac{6p - 7\alpha_s  - p\alpha_s - 6}{2(\delta+8)}}\,\varepsilon^{p-1}_{e,-1}\,\varepsilon^{\frac{p+1}{4}}_{B,-4}\,n_{-4}^{\frac{5\alpha_s+6+\alpha_s p-2p }{4(\alpha_s+6)}}\,D^{-2}_{26.1}\cr
&&\hspace{4.5cm}\times \,\tilde{E}^{\frac{2(p+3)}{\alpha_s+6}}_{51}\, 
\eary
}
and {\small $F_{\rm nu, j}=\{6.1\times \,10^{-3}$, $4.0\times \,10^{-3}$, $6.9\times \,10^{-6}$, $1.1\times \,10^{-7}\}$} mJy for {\small $\epsilon_{\rm \gamma}= \{ 3\, {\rm GHz}$,  $6\, {\rm GHz}$,  $1\, {\rm eV}$,  $1\, {\rm keV}\}$}, respectively.  It is worth noting that for $\alpha_s=0$, the flux $F_{\rm \nu, dec}\propto t^{-p}$ derived in \cite{1999ApJ...519L..17S} is recovered .\\     
Therefore,  the flux to be used  to model the X-ray, optical and radio data can be summarized as
\begin{equation}\label{flux}
F_{\nu}= \left\{
\begin{array} {ll} 
F_{\rm \nu, inc},                            &      \mathrm{if \quad} t<t_{\rm peak}, \\ 
F_{\rm \nu, dec},   &      \mathrm{if \quad} t > t_{\rm peak},
\end{array} \right.
\end{equation}
where  
\be
t_{\rm peak}\simeq86.6\, {\rm day}\,\, {\rm k}\,\left(\frac{1+z}{0.01}\right)\,{\rm n}^{-1/3}_0\,E^{1/3}_{51}\,\Delta \theta_{20^\circ}^{-\frac{\alpha_s+6}{3}}\,,
\ee
where the parameter ${\rm k}$ differs from one model to another, and is introduced to correlate the times of peak flux and the jet break through the viewing and the opening angles \citep{2002ApJ...579..699N, 2002ApJ...570L..61G}. \\
To find these  values,  we again perform the Bayesian statistical method of Markov-Chain Monte Carlo (MCMC) simulations. In this case, our model is fully described by a set of eight parameters, $\Xi_{\rm fow}$ = \{n, $\tilde{E}$, k, $\Delta\theta$, $p$, $\alpha_s$, $\varepsilon_{B,f}$, $\varepsilon_e \}$, with an extra parameter of $\sigma$ for the likelihood of the MCMC.   For this MCMC run, we utilised 14000 steps with 7000 tuning steps, which were discarded after tuning. The parameters $\Delta\theta$  and $p$ are given uniform distributions, while the remaining parameters ${\rm n}$, $\epsilon_{\rm B,f}$, $\epsilon_e$, ${\rm k}$, $\alpha_s$ and $\tilde{E}$ are given normal distributions.   Output is again given by means of a Corner Plot \citep{2016JOSS....1...24F} on which the diagonal is a one-dimensional projection of the posterior probability distribution function and the off-diagonal plots are the bi-dimensional projections.\\
Figures \ref{fig3:param_late_3R},  \ref{fig3:param_late_6R} and \ref{fig3:param_late_X} show the corner plots for radio wavelengths (3 GHz and 6 GHz) and X-rays, respectively, obtained from the MCMC simulation for each parameter using our model (eqs. \ref{flux}).   The best-fit values in these figures are shown in green color.  The median of the posterior distributions, alongside with the symmetrical 35\% quantiles, are  reported in Table \ref{table5:param_late}. 
\subsection{Analysis and implications}
\subsubsection{The magnetic microphysical parameters}
\cite{1995ApJ...455L.143S} derived the hydrodynamic timescales of the reverse shock for a non-magnetized GRB jet.  They found that in absence of magnetization, the crossing time becomes $t_{\rm cr}\simeq \frac{T_{90}}{2}$.  The  hydrodynamic timescales of the reverse shock powered by a  magnetized outflow were  investigated by \cite{2004A&A...424..477F}, \cite{2005ApJ...628..315Z} and \cite{2009A&A...494..879M, 2010MNRAS.407.2501M}.    Authors reported that  general characteristics in the reverse shock vary according to the degree of magnetization in the jet. For instance,  when the jet was moderately magnetized with a magnetization parameter in the range of $0.1\lesssim\sigma\lesssim 1$, then the magnetic microphysical parameter would vary between $0.1 \lesssim \varepsilon_{\rm B,r}\lesssim0.2$ and the width of the peak generated by the reverse shock becomes narrower  and more prominent, between $\frac{T_{90}}{2}  \lesssim t_{\rm cr}\lesssim \frac{T_{90}}{5}$.   This result agrees with  the value of the magnetic microphysical parameter found  after describing the GBM data and the duration  of the bright peak ($\sim$ 0.4 s) observed  in the GBM light curve.  If the relativistic jet  would have had high magnetization ($\sigma\gg$1) when it crosses the reverse shock,  relativistic particles  would  be poorly  accelerated and the emission drastically decreased \citep{2011ApJ...726...75S,2005ApJ...628..315Z}.  Therefore, a moderate  magnetization ($\sigma\lesssim1$) is required in order to interpret the GBM  bright peak in the reverse shock framework \citep{2003ApJ...595..950Z,2003MNRAS.346..905K,2004A&A...424..477F}.\\
The values of the magnetic microphysical parameters (see Tables \ref{table3:param_early} and  \ref{table5:param_late}) indicate that the magnetic field ratio in the forward- and reverse-shock region is $\sim 40$. Similarly, the values found also illustrate that synchrotron flux  is  $\sim2.5\times 10^3$ times  stronger in the reverse than the forward shock, so there are much more photons available to be scattered via inverse Compton in the reverse shock.  It suggests that the outflow carried a significant magnetic field as reported in sGRB 090510 \citep{2016ApJ...831...22F}.\\
Taking into consideration the typical initial value of the fireball radius ($r_i\sim 10^{6.5}\,{\rm cm}$;  \cite{2004ApJ...608L...5L,2005ApJ...632..421L, 2007PhR...442..166N}),  the kinetic equivalent energy and the magnetic microphysical parameter (see Table  \ref{table3:param_early}),   the magnetic field at the base of the jet is roughly estimated as $B\approx\sqrt{8\varepsilon_{B,i}E_{\gamma,iso}/r^3_i}\approx 10^{15}\, {\rm G}$.   The strength of the magnetic field is three-four orders of magnitude higher than usual strength in a NS $\sim 10^{12}\, {\rm G}$. Here, $\varepsilon_{\rm B,i}\approx \varepsilon_{\rm B,r}$ is the initial fraction of total energy given to magnetic field.  It shows that  GRB 170817A demands more magnetic fields at the base of the jet, thus indicating that the progenitor is entrained with strong magnetic fields.\\
\cite{2016ApJ...816L..30J}  laid out  relativistic and axisymmetric hydrodynamic simulations of black hole-torus system as remnants of a binary NS merger. They showed that thermal energy via annihilation of neutrinos and antineutrinos abundantly emitted by the hot accretion disk is not long and strong enough for the outflows to break out from the neutrino wind, thus concluding that the neutrino annihilation alone could not power sGRBs from binary NS mergers. Therefore, the energy requirements favor magnetic fields as the responsible mechanism so that the outflow  breaks out.  Some authors have presented simulations based on general relativistic magnetohydrodynamics to follow the evolution of the magnetic fields in the binary NS merger \citep{2006Sci...312..719P, 2013ApJ...769L..29Z, 2017PhRvD..95f3016C}.   All models proposed show an amplification of magnetic field up to three orders of magnitude or more.  The rapid growth of this field is attributed to the  Kevin-Helmholtz instabilities and turbulent and/or differential rotation.   Therefore, the most natural process associated with the magnetization of outflow  is the magnetic field amplification during the binary NS merger which is entrained by outflow.\\
%
%
%
\subsubsection{Other parameters}
\begin{enumerate}
\item  The values of the external medium densities required to model the  $\gamma$-ray GBM data $\sim$ 1.7 s after the merger (see Table \ref{table3:param_early}) and the X-ray, optical and radio data (see Table \ref{table5:param_late})  are quite different, indicating that the $\gamma$-ray emission and the afterglow occurred in different regions.   It suggests that the external density distribution could be stratified as proposed in sGRBs \citep{2009arXiv0904.1768P}.   Binary NS mergers are thought to be potential candidates to eject significant masses with distinct velocities and densities. The ejected masses with densities larger than low ISM are ejected at sub-relativistic velocities. In principle,  the ultra-relativistic jet coming out from the progenitor could interact with these  dense material producing an afterglow \citep{2015MNRAS.450.1430H, 2013ApJ...778L..16H, 2014MNRAS.437L...6K, 2001ApJ...551..946T, 2018arXiv180302978F}. Moreover, fits to the multiwavelength afterglow have been suggested that the circumburst medium close to the progenitor could be dense and be formed by gaseous environments rather than  the low ISM \citep{2007ApJ...670.1254B, 2009ApJ...701..824N, 2006MNRAS.367L..42P, 2007PhR...442..166N, 2009arXiv0904.1768P}. The low value of the ISM would confirm that sGRBs explode in lower-density environments. 
\item The values of the electron spectral indexes for the $\gamma$-ray flux and the X-ray, optical and radio fluxes  are equal.   These  spectral indexes correspond to  the typical values reported in external shocks $2.2\leq p \leq 2.6$  \cite[e.g., see;][]{2015PhR...561....1K}.   It suggests that the GBM $\gamma$-ray flux  could have been originated in external shocks.  Similar results have been found in several bursts that have exhibited early sub-GeV $\gamma$-ray and optical peaks together with temporarily extended multiwavelength emissions \citep{2007ApJ...655..973K, 2007ApJ...655..391K, 2015ApJ...804..105F, 2016ApJ...818..190F, 2017ApJ...848...15F}.   
\item Two scenarios are discussed in order to explain GRB 170817A \citep[e.g.,][]{2017ApJ...848L..34M}: a low-luminosity sGRB and a typical sGRB viewed off-axis. Whereas a low-luminosity sGRB could be produced by a mildly relativistic outflow   \citep{2003MNRAS.343L..36R, 2002MNRAS.336L...7R, 2014ApJ...784L..28N}, a typical sGRB is generated by a relativistic jet \citep{2005ApJ...625L..91R,2014ApJ...788L...8M}. In both cases a relativistic jet is invoked, however in the first case the jet is choked by the wind expelled from the hyper massive neutron star (HMNS), thus giving rise to a low-luminosity sGRB with $E_{\rm \gamma, iso}\simeq 10^{46}$ - $10^{47}$ erg. Considering the values of the equivalent energy we estimate for this event (see Tables \ref{table3:param_early} and  \ref{table5:param_late}), we suggest that the most likely scenario for GRB 170817A is that of a jet that successfully breaks out from the wind and it is viewed off-axis. This result agrees with the recent work by \cite{2018arXiv180609693M}, where authors present Very Long Baseline Interferometry (VLBI) observations that show superluminal motion and support the successful breakout of the jet.
\item The collimation of ejecta has relevant implications in GRBs. For instance,  the energy scale, the energy extraction mechanism and the event rate. For sGRBs, there are only a few observations of jet breaks despite serious effort.  Based on the breaks detected in the afterglow emission,  \cite{2014ARA&A..52...43B} showed a distribution of jet opening angles for sGRBs with a mean around $\theta_j\sim\langle 5^\circ\rangle$. Recently, a similar value of opening angle was obtained after modeling the afterglow in GRB 170817A \citep[i.e. see;][]{2018arXiv180106516T, 2017arXiv171006421G}. Taking into account the value of $\theta_j= 5^\circ$, the viewing angle for GRB 170817A would be $\theta_{\rm obs}=20^\circ$, which is in the range reported for this burst \citep{2017ApJ...848L..20M, 2017arXiv171006421G, 2018arXiv180103531M, 2018arXiv180106516T, 2018arXiv180609693M}. 
\item Considering the values reported together with eqs. (\ref{Eoff}) and (\ref{newLorentz}), the bulk Lorentz factor is $\Gamma\simeq8.7$ and the equivalent kinetic energy is  $E_{\rm k}\simeq3.3\times 10^{47}\,{\rm erg}$.  Comparing with the observed isotropic energy $E_{\rm \gamma, iso}\simeq5\times 10^{46}$ erg, the corresponding efficiency becomes $\eta\simeq 15\%$, which lies in the typical range reported for afterglows \citep[e.g. see;][]{2015PhR...561....1K}. 
\end{enumerate}
\section{Conclusions}
We have analyzed the non-thermal  ($\gamma$-ray, X-ray, optical and radio) observations of GRB 170817A/GW170817.   The X-ray, optical and radio data were consistent  with the synchrotron forward-shock model  when  the jet is viewed off-axis and the matter in the outflow is parametrized through a power law velocity distribution.  The origin of the $\gamma$-ray peak was discussed in terms of internal and external shocks. The analysis performed favors to a SSC reverse-shock model  in the fast-cooling regime observed at high latitudes.
%
%
%
The fit of the  $\gamma$-ray GBM data with SSC model suggests that:

\begin{itemize}

\item The circumburst medium close to the progenitor is much denser than the low ISM obtained after modeling the X-ray, optical and radio data. One possible explanation suggests that the external density distribution could be stratified as proposed in sGRBs \citep{2009arXiv0904.1768P, 2015MNRAS.450.1430H, 2013ApJ...778L..16H, 2014MNRAS.437L...6K, 2001ApJ...551..946T, 2018arXiv180302978F};  the circumburst medium close to the progenitor could be formed by gaseous environments and/or dense ejected masses rather than a very low density medium of the host galaxy.  It suggests that the afterglow and $\gamma$-ray emission occurred in different regions.

\item The value of the electron spectral index illustrates that this component could have been originated at the external shocks.  Similar discussions have been previously reported  around the temporarily extended Fermi-LAT components \citep{2007ApJ...655..973K, 2007ApJ...655..391K, 2015ApJ...804..105F, 2016ApJ...818..190F, 2017ApJ...848...15F}.

\item  The value of the magnetic microphysical parameter obtained agrees with  the temporal properties exhibited by this burst  and also indicates that the strength of the magnetic field is three-four orders of magnitude higher than usual strength in a NS $\sim 10^{12}\, {\rm G}$.    By comparing  the magnetic microphysical parameters obtained for $\gamma$-ray flux with the X-ray, optical and radio observations is shown that the magnetic field in the  reverse-shock region would be $\sim 40$ times higher than the forward shock.   It suggests that the outflow carried a significant magnetic field as reported in sGRB 090510 \citep{2016ApJ...831...22F}.

\end{itemize}
The value of the equivalent kinetic energy agrees with simulations performed around the necessary conditions for sGRB production in binary NS mergers \citep{2014ApJ...788L...8M}. It suggests  the scenario of the collapse to a black hole with the formation of a typical off-axis sGRB favours on that where the wind expelled from HMNS hampers the forward movement of the on-axis jet.\\
Since GRB 170817A was the closest sGRB with measured redshift,  it  was proposed as potential target for neutrino observation. However,  the  Antares, IceCube and Auger observatories reported a null result based on a search  during the prompt phase and afterglow \citep{2017arXiv171005839A, 2017GCN.21568....1N}.   As showed in previous works  \citep[see i.e.,][]{2013ApJ...772L...4G, 2017ApJ...848...15F},  the lack of energetic neutrinos around GRB 170817A could be related with the degree of the ejecta magnetization which hinders efficiently particle acceleration \citep{2011ApJ...726...75S}. \\
%
One of the most energetic short bursts,  GRB 090510 located at z=0.903, was detected by Fermi and Swift satellites \citep{2010ApJ...709L.146D}.  This sGRB seen on-axis exhibited a short-lasting peak at the end of the prompt phase ($T_{90}=0.3$ s) and a temporally extended component lasting hundreds of seconds.   In addition,  Ultra Violet and Optical Telescope (UVOT) on board of the Swift satellite started collecting optical data at 97 s after the initial trigger \citep{2009GCN..9342....1K}.  The optical afterglow emission was described by a broken power law with the best-fit parameters:  an early decay slope of $-0.50^{+0.11}_{-0.13}$, a break time of $1.58^{+0.46}_{-0.37}\times 10^3$ s, a late decay slope of $1.13^{+0.11}_{-0.13}$ and density flux of $\sim10^{-13}\,{\rm erg\,cm^{-2}\,s^{-1}}$ at one day after the trigger. \cite{2016ApJ...831...22F} used an early-afterglow model to interpret the multiwavelength light curve observations. In particular,  SSC emission from the reverse shock was consistent with the bright LAT peak provided that the progenitor was endowed with strong magnetic fields, thus associating this progenitor with a binary NS merger. The optical light curve was described by synchrotron forward-shock emission in the slow cooling regime before and after the break time.  A similar analysis for an off-axis emission was done in this paper for GRB 170817A.  The bright $\gamma$-ray peak was consistent  with SSC radiation in the fast-cooling regime and the multiwavelength afterglow with synchrotron emission in the slow-cooling regime at different regions.  Therefore,  we argue that an amplification process related to the  binary NS merger in GRB 170817A was present.   This burst did not display high-energy photons ($> 100\,{\rm MeV}$) probably due to the high charged particle background in the burst region \citep{2017GCN.21534....1N}, the off-axis emission and low isotropic energy.\\ 
%
Gravitational wave observations from a  binary NS merger associated with this GRB  event \citep{2017GCN.21520....1V, 2017arXiv171005446G} cast the compact merger scenario in new light.  Similar analysis to the one presented here on future short GRBs can shed light on the nature of the progenitors, evolution of magnetic field and optical counterpart  addressing the short GRB-gravitational wave association.\\
\\
\vspace{1cm}
%
%
\section*{Acknowledgements}
We thank the referee for a critical reading  and significant suggestions that helped improve this  manuscript.   We also thank E. Ramirez-Ruiz  for useful discussions.  NF  acknowledges  financial  support from UNAM-DGAPA-PAPIIT through grants IA102917 and IA102019. FDC thanks the UNAM-PAPIIT grants IN117917 and support from the Miztli-UNAM supercomputer (project LANCAD-UNAM-DGTIC-281).  PV  thanks  Fermi  grants NNM11AA01A and 80NSSC17K0750, and  partial support from OTKA NN 111016 grant. RBD acknowledges support from the National Science Foundation under Grant 1816694. ACCDESP acknowledges that this study was financed in part by the Coordenação de Aperfeiçoamento de Pessoal de Nível Superior - Brasil (CAPES) - Finance Code 001 and also thanks the Professor Dr. C. G. Bernal for tutoring and useful discussions. 
%
%

%
\clearpage
\begin{table}
\centering
\caption{Spectral analysis with GBM data}\label{table1:gbm_analysis}
\begin{tabular}{ l c c c c c}
 \hline
 \scriptsize{Time Interval (s)$^a$} & \scriptsize{Model $^b$} & \scriptsize{$\beta$} &\scriptsize{$E_{\rm pk}$ (keV)}& kT (keV)  &\scriptsize{C-Stat/dof}  \\
 \hline 
 \hline 
\scriptsize{[-0.320 , 0.320]}  & \scriptsize{CPL}  &  \scriptsize{-1.016$\pm$0.293} & \scriptsize{338.3$\pm$229} &  -     & \scriptsize{406.64/361}   \\
\scriptsize{[-0.320 , 0.256]}  & \scriptsize{CPL}  &  \scriptsize{-0.955$\pm$0.309} & \scriptsize{331.6$\pm$212} &   -    & \scriptsize{414.34/361}   \\
\scriptsize{\,[0.256 , 0.320]}   &  \scriptsize{PL}  &  \scriptsize{-1.749$\pm$0.434} & \scriptsize{-} &  - &\scriptsize{296.47/362}   \\
\scriptsize{\,[0.320 , 0.448]}   &  \scriptsize{PL}  &  \scriptsize{-2.150$\pm$0.472} & \scriptsize{-} &  - &\scriptsize{341.51/362}   \\
\scriptsize{\,[0.512 , 1024]}   &  \scriptsize{BB}  &  \scriptsize{-} & \scriptsize{-} &  \scriptsize{13.84 $\pm$ 4.67} &\scriptsize{446.40/362}   \\
\scriptsize{\,[1.024 , 1.536]}   &  \scriptsize{BB}  &  \scriptsize{-} & \scriptsize{-} & \scriptsize{11.78 $\pm$  2.41}  &\scriptsize{399.07/362}   \\
\scriptsize{\,[1.536 , 2.048]}   &  \scriptsize{BB}  &  \scriptsize{-} & \scriptsize{-} & \scriptsize{9.480 $\pm$ 1.61}  &\scriptsize{416.05/362}   \\

\hline
\end{tabular}
\begin{flushleft}
\scriptsize{
$^a$ Time interval is given to GBM trigger.\\
$^b$ CPL = Comptonized function.   PL = Simple power law function.  BB= Black-body funcion\\
}
\end{flushleft}
\end{table}
\begin{table}
\centering
\caption{Fitted values of the $\gamma$-ray data.\\   The chi-square minimizations ($\chi^2$ / n.d.f.)  are reported in parenthesis}\label{table2: fit_gbm}
\begin{tabular}{ l c c}
\hline
\hline

{\large{$\gamma$-ray flux}} 	& 			 				                         &	  		 	        	   		 \\ 
\hline \hline
\\
\small{Decay slope}	&  \small{$\alpha_{\gamma}$\hspace{0.5cm}}	 	&		\small{$2.85\pm0.35$}  {\footnotesize (4.27/4)}  \\
\small{Starting time (s)} &\small{$t_0$} \hspace{0.5cm}	 	                         &               \small{$2.0\pm0.1$}  {\footnotesize (4.27/4)}   \\
\small{Flux rise timescale (s)} &\small{$\tau$} \hspace{0.5cm}	 	                         &               \small{$0.4\pm0.1$}  {\footnotesize (4.27/4)}   \\
\hline
\end{tabular}
\end{table}
\begin{table}
\centering \renewcommand{\arraystretch}{2}\addtolength{\tabcolsep}{-4pt}
\caption{The median and symmetrical quantiles (0.15, 0.5, 0.85) are reported  after describing the $\gamma$-ray GBM peak with our model.}\label{table3:param_early}
\begin{tabular}{ l c c}
\hline
\hline

{\large Parameters} 	& 	{\large Median} 			 				                                  \\ 
\hline \hline
\small{$\tilde{E}\, (10^{51}\,\,{\rm erg})$}	&   \small{$0.83^{+1.19}_{-0.54}$}	 			  \\
\small{${\rm n}\,\, ({\rm cm^{-3}}$ ) }	        	&   \small{$1.01^{+0.29}_{-0.29}\,$}			 	  \\
\small{$\Gamma_r$}	                                 	&   \small{$24.94^{+4.93}_{-4.84}$}			        	  \\
\small{${\rm p}$ }	                                  &   \small{$2.20^{+0.06}_{-0.06}$}	 			   \\
\small{$\Delta \theta$\,({\rm deg})}		&   \small{$15.01^{+0.68}_{-0.68}$}	        			   \\
\small{$\varepsilon_e\,\,(10^{-1})$}  		&   \small{$3.17^{+0.83}_{-1.16}$}	 	                    \\
\small{$\varepsilon_{B,r}\,\,(10^{-1})$}			&   \small{$1.80^{+1.15}_{-0.78}$}	 	                     \\
\hline
\end{tabular}
\end{table}

\begin{table}
\centering
\caption{Fitted values of the X-ray, optical and radio data.\\   The chi-square minimizations ($\chi^2$ / n.d.f.)  are reported in parenthesis}\label{table4:fit_root}
\begin{tabular}{ l c c c}

\\
\hline
\\
{\normalsize  \bf Light Curve}\\		                           
\hline \hline
\small{\bf{X-ray flux}} 	& 			 				                         &	  		 	        	   		 \\ 
\hline \hline  
\small{Rising slope}	&  \small{$\alpha_{X}$\hspace{0.5cm}}	 	&		\small{$0.76\pm0.18$}  {\footnotesize (0.45/4)}  \\
%
\\
\hline \hline
\small{\bf{Optical flux}}		         & 		                                          &                                        \\
\hline \hline
\small{Rising slope} &\small{$\alpha_{\rm opt}$\hspace{0.5cm}}			 & \hspace{1cm}\small{$-$} \hspace{2.1cm}    \\
%
\\
\hline\hline
\small{\bf{Radio flux}}		         & 		                                          &                                        \\
\hline \hline
\small{3 GHz} & &\\
\cline{1-1}
\small{Rising slope } &\small{$\alpha_{\rm 3GHz}$\hspace{0.5cm}}			 & \small{$0.85\pm0.12$}  {\footnotesize(1.67 / 3)} \\
%
\\
\hline
\small{6 GHz} & & \\
\cline{1-1}
\small{Rising slope} &\small{$\alpha_{\rm 6GHz}$\hspace{0.5cm}}			 & \small{$0.75\pm0.19$}  {\footnotesize(11.16 / 6)}\\
\\
\hline
\\
{\normalsize  \bf Spectral Energy Distribution}\\	
 \hline\hline
\small{Spectral slope (16 $\pm$ 2 d)}                            &\small{$\beta_{\rm 16d}$}		&				\small{\hspace{0.6cm}} & \small{$-0.59\pm0.11$} {\footnotesize (3.796 / 7)}	\\
\small{Spectral slope (110 $\pm$ 5 d)}                          &\small{$\beta_{\rm 110d}$}		&				\small{\hspace{0.6cm}} &\small{$-0.58\pm0.15$} {\footnotesize(19.19 / 20)}	\\
\small{Spectral slope (145 $\pm$ 20 d)}                        &\small{$\beta_{\rm 145d}$}		&				\small{\hspace{0.6cm}} &\small{$-0.59\pm0.15$} {\footnotesize(19.19 / 20)}	\\
\\
\hline

\end{tabular}
\end{table}

\begin{table}
\centering \renewcommand{\arraystretch}{2}\addtolength{\tabcolsep}{3pt}
\caption{The median and symmetrical quantiles (0.15, 0.5, 0.85), truncated at the second decimal,  are reported after describing the X-rays and radio wavelengths at 3 and 6 GHz with our model.}\label{table5:param_late}
\begin{tabular}{ l  l  l  l c }
\hline
\hline

{\large   Parameters}	& 		& {\large  Median}  & 		 				                           		 	        	   		 \\ 
                          	& 	{\normalsize Radio (3 GHz)} 	&  {\normalsize Radio (6 GHz)} & 		{\normalsize X-ray (1 keV)}			                           		 	        	   		 \\ 

\hline \hline
\\
\small{$\tilde{E}\, (10^{51}\,{\rm erg})$}	\hspace{1cm}&   \small{$0.700^{+0.010}_{-0.010}$}	 \hspace{0.7cm}	&  \small{$0.700^{+0.010}_{-0.010}$}	\hspace{0.7cm} &  \small{$0.701^{+0.010}_{-0.010}$}	 \\
\small{${\rm n}\,\, (10^{-4}\,{\rm cm^{-3}}$ ) }	\hspace{1cm}	&  \small{$1.010^{+0.010}_{-0.010}$} 	 \hspace{0.7cm}&  \small{$1.020^{+0.010}_{-0.010}$} 	\hspace{0.7cm}&  \small{$1.008^{+0.010}_{-0.010}$}	 	                           \\
\small{${\rm p}$}	\hspace{1cm}&  \small{$2.210^{+0.010}_{-0.010}$}  \hspace{0.7cm}&  \small{$2.210^{+0.010}_{-0.010}$} \hspace{0.7cm}&  \small{$2.230^{+0.010}_{-0.010}$}	 	                                \\
\small{$\Delta \theta$\,({\rm deg})}	\hspace{1cm}&  \small{$15.001^{+0.133}_{-0.136}$}	 \hspace{0.7cm}&  \small{$15.001^{+0.137}_{-0.137}$}	\hspace{0.7cm} &  \small{$15.001^{+0.133}_{-0.136}$}	 	                 \\
\small{$\varepsilon_e\,\,(10^{-1})$}   \hspace{1cm}&\small{$2.500^{+0.010}_{-0.010}$}  \hspace{0.7cm}&\small{$2.500^{+0.010}_{-0.010}$} \hspace{0.7cm}&\small{$2.498^{+0.010}_{-0.010}$}	 	                           \\
\small{$\varepsilon_{B,f}\,\,(10^{-4})$}	 \hspace{1cm}&\small{$1.010^{+0.010}_{-0.010}$}   \hspace{0.7cm}&\small{$1.100^{+0.010}_{-0.010}$} \hspace{0.7cm} &\small{$0.997^{+0.010}_{-0.010}$}	 	    \\
\small{${\rm k}$}  \hspace{1cm}&\small{$3.000^{+0.010}_{-0.010}$}   \hspace{0.7cm}&\small{$3.000^{+0.010}_{-0.010}$} \hspace{0.7cm} &\small{$2.998^{+0.010}_{-0.010}$}	 	    \\
\small{${\alpha_s}$}  \hspace{1cm}&\small{$1.105^{+0.010}_{-0.010}$}   \hspace{0.7cm}&\small{$1.095^{+0.010}_{-0.010}$} \hspace{0.7cm} &\small{$1.115^{+0.004}_{-0.007}$}	 	    \\

\hline
\end{tabular}
\end{table}


\begin{figure}
{ \centering
\resizebox*{0.5\textwidth}{0.4\textheight}
{\includegraphics{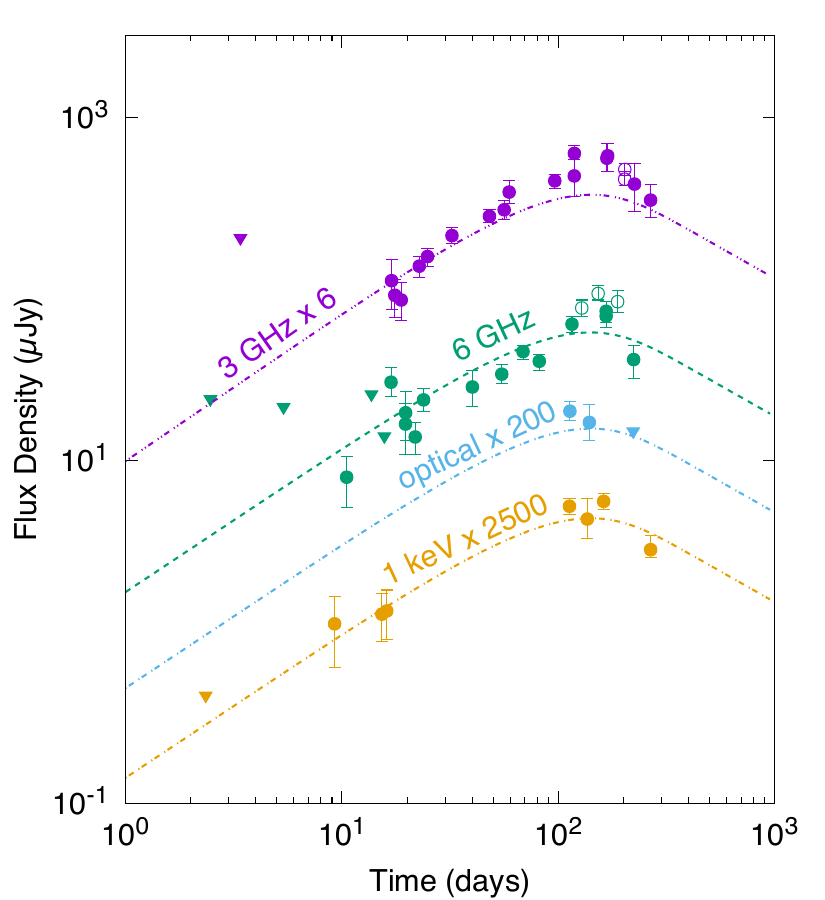}}
\resizebox*{0.5\textwidth}{0.4\textheight}
{\includegraphics{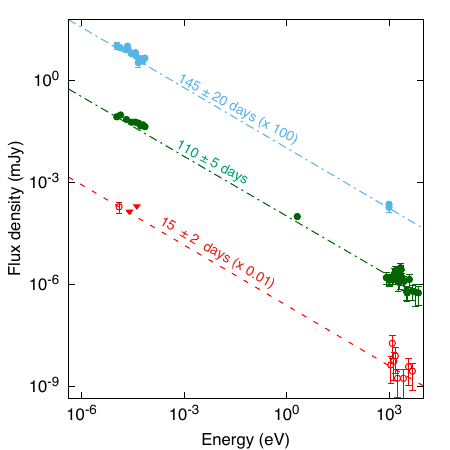}}
}
\caption{Right: SEDs of the X-ray, optical and radio afterglow observations at 15 $\pm$ 2 (red),  110 $\pm$ 5 (green) and 145 $\pm$ 20 (blue) days.  Left: Light curves of X-ray at 1 keV  \citep[gold;][]{2017Natur.547..425T, 2017ApJ...848L..20M, 2017ATel11037....1M, 2018ATel11242....1H,  troja2017a, 2018arXiv180106516T,2018arXiv180103531M},  optical  \citep[blue;][]{2018arXiv180103531M},  and radio  at 3 and 6 GHz \citep[magenta and green;][]{2017Natur.547..425T, 2017arXiv171005435H, 2017arXiv171111573M, 2017ApJ...848L..21A}   bands.  The values that describe both the SED and the light curves are reported in Table \ref{table5:param_late}. }
\label{fig4:afterglow}
\end{figure}

\end{document}